\documentclass[%
 amsmath,amssymb,
 aps,
]{revtex4-2}

\usepackage{graphicx}
\usepackage{dcolumn}
\usepackage{bm}
\usepackage{amssymb,amsmath}
\usepackage{subfig}
\usepackage{graphicx}
\usepackage{braket}
\usepackage{physics}
\usepackage{color}
\usepackage{hyperref}
\usepackage{booktabs}
\usepackage{tabularx}
\usepackage{listings}
\usepackage[utf8]{inputenc}
\usepackage{dsfont}
\usepackage{xcolor}

\begin{document}

\preprint{Preprint}

\title{Tripartite entanglement: Foundations and applications}

\author{M\'{a}rcio M. Cunha $^{1}$, Alejandro Fonseca $^{2}$ and Edilberto O. Silva $^{1,\ddagger}$*}

\address{$^{1}$ \quad Departamento de F\'{i}sica, Universidade Federal do Maranh\~{a}o, S\~{a}o Lu\'{i}s, MA 65085-580, Brazil; marciomc05@gmail.com\\
$^{2}$ \quad Departamento de F\'{i}sica, Universidade Federal de Pernambuco, 50670-901, Recife, Pernambuco, Brazil; alejofonseca4@gmail.com\\
$^{\dagger}$ Correspondence:  edilbertoo@gmail.com}




\date{\today}
\begin{abstract}
We review some current ideas about tripartite entanglement, the case representing the next level of complexity beyond the simplest one (though far from trivial), namely the bipartite. This kind of entanglement has an essential role in the understanding of foundations of quantum mechanics. Also, it allows several applications in the fields of quantum information processing and quantum computing. In this paper, we make a revision about the main foundational aspects of tripartite entanglement and we discuss the possibility of using it as a resource to execute quantum protocols. We present some examples of quantum protocols in detail.\\
\textbf{$\rightarrow$} \textit{This version contains an Erratum:} $\rightarrow$ We fix some mistakes presented in the previous version.
\end{abstract}             

\maketitle

\section{Introduction}

Quantum entanglement is one of most astonishing aspects of Quantum Mechanics, initially due to the deep implications in the context of the theory itself and more recently because of the large amount of applications within the emerging fields of Quantum Information \cite{Nielsen} and Quantum Computation\footnote{Note though that entanglement is not the only non-classical resource useful for computation.}.

The most well-known type of entanglement involves two parts sharing two qubits, namely, the EPR or sometimes called Bell states \cite{barnett2009quantum}. Nevertheless, note that two parts may also share an entangled state in larger dimensions, like qutrits \cite{PhysRevLett.123.070505}.

Besides the relevance of bipartite entanglement in the understanding of quantum foundations and quantum information science, the search of entangled states involving more than two qubits is also desirable, because it opens new possibilities on the fundamental aspects of the theory itself and also in development of new protocols in Quantum Information \cite{PhysRevA.74.032324}.
Also, dealing with genuine multipartite entanglement can provide several advantages in comparison with bipartite entanglement \cite{de2011quantum}. It is possible to establish quantum networks with multi-users, execute quantum computation by using cluster states \cite{PhysRevLett.86.5188,PhysRevA.96.052314} and also perform measurement-based quantum computing \cite{briegel2009measurement}.
These entangled states could be used, for instance, as a quantum channel to establish quantum communication between several separated locations.

The simplest case of multipartite entanglement it is tripartite entanglement, that involves three-parts. These type of entanglement have an essential role in the development of aspects like quantum non-locality and a large number of applications in quantum information protocols. In this review, we explore the main features of tripartite entanglement, from the most fundamental aspects to applications.

The paper is organized in the following way: In section 2, we consider a brief overview of bipartite entanglement. In section 3, we make a discussion about tripartite entanglement and define the corresponding classes of entanglement for qubits.
Then, we consider some aspects of quantum non-locality and tripartite entanglement in section 4. In section 5, we consider several examples of quantum information protocols employing tripartite entanglement. We review the main aspects of these applications, illustrating the corresponding schemes.
In section 6, we list several contributions to the production of tripartite entanglement in the literature. Then, we proceed considering the detection and characterization of tripartite entanglement in section 7. In section 8, we consider the topic of Remote Preparation of quantum states. In section 9, we include aspects of tripartite entanglement involving continuous variables. The effects of noisy environments it is considered in section 10. Finally, we make our conclusions in section 11.

\section{Overview on Bipartite Entanglement}

For the sake of completeness, let us quickly review some important aspects of entanglement between two parts\footnote{It is important to keep in mind that \textit{bipartite entanglement} does not necessarily imply two spatially separated particles, instead it can be generated between different degrees of freedom in a single part \cite{Gabriel2011}.}. We start by describing the case of pure states and then we cite some of the most important entangled mixed states.

\subsection{Pure states}

Due to the Schmidt decomposition, any quantum state shared between two parts, say Alice and Bob as usual, may be written as $\ket{\psi}=\sum_{j=0}^{d-1}\sqrt{\lambda_j}\ket{\phi_A}_j\otimes\ket{\varphi_B}_j=\sum_{j=0}^{d-1}\sqrt{\lambda_j}\ket{j,j}$, with Schmidt coefficients $\lambda_j\in\mathbb{R}$, satisfying $\sum_{j=0}^{d-1}\lambda_j=1$ and $d=\min(\dim{\mathcal{H}_{A},\dim{\mathcal{H}_{B})}}$, where $\mathcal{H}_{k}$ is the Hilbert space associated to the $k-$th part. In particular, it is possible to say that $\ket{\psi}$ is entangled whenever there exist more than one non-null Schmidt coefficients. Furthermore, we can define a basis for the Hilbert space associated to both parts $\mathcal{H}_{AB}=\mathcal{H}_{A}\otimes\mathcal{H}_{B}$, composed by $d^2$ elements given by
\begin{equation}
\ket{\phi_{mn}^{(d)}}=\sum_{k=0}^{d-1}\omega_d^{mk}\beta_{km}\ket{k,k\oplus n},\;\;\;\;\;m,n=0,\dots, d-1,
\end{equation}
where $\omega_d=\exp(2\pi i/d)$ is the primitive $d$-th root of unity, the symbol ``$\oplus$" denotes sum modulo $d$ and the $\beta_{km}$ coefficients control how entangled the basis is. In particular, for $d=2$ the basis can be parametrized as  \cite{PhysRevA.92.012338}
\begin{equation*}
\ket{\phi_{00}^{(2)}}=\cos\theta\ket{00} +\sin\theta\ket{11}, \;\;\;\;\; \ket{\phi_{10}^{(2)}}=\sin\theta\ket{00} -\cos\theta\ket{11},
\end{equation*}
\begin{equation*}
\ket{\phi_{01}^{(2)}}=\cos\theta\ket{01} +\sin\theta\ket{10}, \;\;\;\;\;
\ket{\phi_{11}^{(2)}}=\sin\theta\ket{01} -\cos\theta\ket{10},
\end{equation*}
with $0\leq \theta \leq \pi/2$. Moreover any element of the basis with $\beta_{km}=1/\sqrt{d}$ corresponds to a maximally entangled state. Explicitly, it is known as Bell or EPR\footnote{EPR, after Einstein, Podolski and Rosen's seminal paper \cite{PhysRev.47.777})} basis for $d=2$
\begin{equation}
\ket{\phi^{\pm}}=\frac{1}{\sqrt{2}}\left(\ket{00}\pm \ket{11}\right), \hspace{0.4cm} \ket{\psi^{\pm}}=\frac{1}{\sqrt{2}}\left[\ket{01}\pm \ket{10}\right].
\end{equation}
Hereafter $\ket{\phi_{mn}^{(d)}}$ denotes a maximally entangled state, for the sake of simplicity.

An important feature is that any Bell state can be converted into another one by using local unitary transformations and classical communication (hereafter LOCC). Moreover, it has been shown that these states can be used to develop several informational tasks such as superdense coding and quantum teleportation with the highest attainable performance, and also represent a very useful resource in tests to investigate fundamental aspects of the quantum world.

\subsection{Some special families of mixed states}

Suppose Alice and Bob have a source of entangled qudits prepared in a state $\hat{\rho}$. Then, she applies an unitary operation $\hat{U}$ chosen at random and informs Bob to carry out either $\hat{U}$ or $\hat{U}^*$\footnote{The symbol \textquotedblleft $^*$\textquotedblright  indicates complex conjugation of the associated matrix elements.} on his qudit. Regardless the initial state of the system, after many repetitions of the same procedure, the final state shared by Alice and Bob reduces to a Werner or an isotropic state.

\subsubsection{Werner states}
\label{WS}

When Alice and Bob apply local operations $\hat{U}\otimes \hat{U}$, we have \cite{Werner89}
\begin{equation}
\hat{\rho}\rightarrow \int \hat{U}\otimes \hat{U} ~\hat{\rho}~\hat{U}^{\dagger}\otimes \hat{U}^{\dagger} \dd U = \hat{\rho}_W,
\end{equation}
where $\dd U $ is the Haar measure of the unitary group $\textbf{U}(d)$ and $\hat{\rho}_W$ is the Werner state, given by
\begin{equation}
\hat{\rho}_W=(1-p)\frac{2}{d^2+d}\hat{P}^{(+)}+p\frac{2}{d^2-d}\hat{P}^{(-)},
\end{equation}
with $\hat{P}^{(\pm)}=\frac{1}{2}\left(\hat{1}\pm\hat{V}\right)$, where $\hat{1}$ is the identity: $\hat{1}=\sum_{jk=0}^{d-1}\dyad{jk}{jk}$ and $\hat{V}$ is the flip operator: $\hat{V}=\sum_{jk=0}^{d-1}\dyad{jk}{kj}$. It is important to mention that the Werner state $\hat{\rho}_W$ is invariant under $\hat{U}\otimes \hat{U}$ operations. By using the following relations $\hat{P}^{(+)}\hat{P}^{(-)}=0$, $\hat{P}^{(\pm)2}=\hat{P}^{(\pm)}$ and $\tr\hat{P}^{(-)}=\frac{1}{2}(\tr\hat{1}-\tr\hat{V})=\frac{1}{2}(d^2-d)$, it is easy to show that the operation invariant $p$ \footnote{Particularly in the case $d=2$, $p$ denotes the singlet fraction or in other words the degree of similarity of the state before ($\hat{\rho}$) and after ($\hat{W_p}$) twirling operations, to the singlet state $\ket{\phi_{11}^2}$, i.e. $p=\tr(\dyad{\phi_{11}^2}{\phi_{11}^2}\hat{\rho})=\tr(\dyad{\phi_{11}^2}{\phi_{11}^2}\hat{W_p})$.} is equal to $p=\tr (\hat{P}^{(-)}\hat{\rho}_W)$.

\subsubsection{Isotropic states}

In the case of local operations $\hat{U}\otimes\hat{U}^*$, the state $\hat{\rho}$ is transformed as \cite{Horodecki99}
\begin{equation}
\hat{\rho}\rightarrow \int \hat{U}\otimes \hat{U^*} ~\hat{\rho}~\left(\hat{U}\otimes \hat{U^*}\right)^{\dagger} \dd U=\hat{\rho}_f,
\end{equation}
where $\hat{\rho}_f$ is the isotropic state, given by
\begin{equation}
\hat{\rho}_f=\frac{1-f}{d^2-1}\hat{1}+\frac{fd^2-1}{d^2-1}\hat{P}_+,
\label{rhof}
\end{equation}
with $\hat{P}_+=\dyad{\phi_{00}^{(d)}}{\phi_{00}^{(d)}}$. Analogously to the Werner state, the isotropic state is invariant under $\hat{U}\otimes \hat{U}^*$ operations.

By expanding the identity operator in the generalized Bell basis $\hat{1}=\sum_{\mu\nu=0}^{d-1}\dyad{\phi_{\mu\nu}^{(d)}}{\phi_{\mu\nu}^{(d)}}$, the isotropic state takes the following form:
\begin{equation}
\hat{\rho}_f=f\dyad{\phi_{00}^{(d)}}{\phi_{00}^{(d)}}+\frac{1-f}{d^2-1}\sum_{\substack{\mu\nu=0\\ (\mu,\nu)\neq (0,0) }}^{d-1}\dyad{\phi_{\mu\nu}^{(d)}}{\phi_{\mu\nu}^{(d)}}.
\label{rhof2}
\end{equation}
In this expression it is possible to see more clearly that the operation invariant $f$ is equal to: $f=\tr(\dyad{\phi_{00}^{(d)}}{\phi_{00}^{(d)}}\hat{\rho})=\tr(\dyad{\phi_{00}^{(d)}9}{\phi_{00}^{(d)}}\hat{\rho}_f)$.
Due to their properties, these states have been very useful to unveil the relation between the concepts of entanglement and Bell non-locality \cite{Brunner2014,Hirsch2016}. Several proposed generalizations to the multipartite case are exposed later in this review.

\section{Tripartite entanglement}

\begin{figure}[h!]
\centering
\includegraphics[width=7 cm]{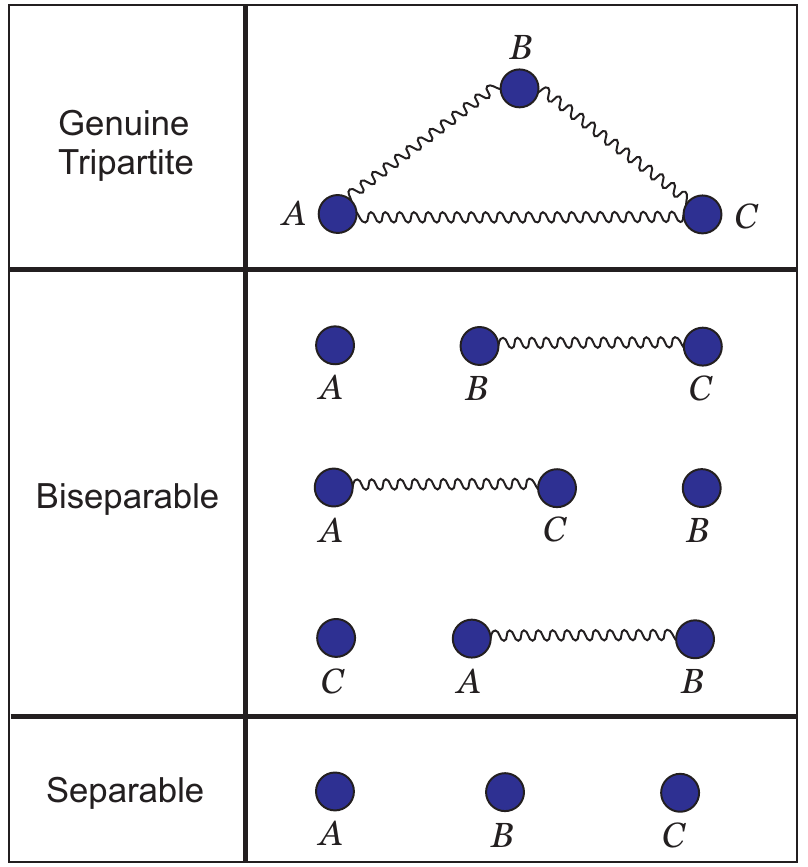}
\caption{All the possibilities of how entangle three qubits. From the top to the bellow: First, Genuine tripartite entanglement, where all qubits are in a entangled state. Second, it is possible to have biseparable states, where two qubits are entangled, and another one it is separated. Third, three qubits in a state full separable.}
\label{ABC}
\end{figure}

\subsection{Genuine tripartite pure states}

When dealing with tripartite quantum systems it is possible to write the associated state $\ket{\Psi}$, in one out of three ways: totally separable $\ket{\Psi}=\ket{\psi_A}\otimes\ket{\psi_B}\otimes\ket{\psi_C}$, biseparable partitions  $\ket{\Psi}=\ket{\psi_j}\otimes\ket{\psi_{kl}}$ ($\{j,k,l\}=\{A,B,C\}$) or genuinely entangled, explained in the next part. In \cite{Zhao2013}, Zhao and collaborators provide a set of methods based on expectation values of Pauli operators to identify the class of a given state.

It is a well known fact that it is possible to transform any state from the Bell basis into anther one by using LOCC only, or to another arbitrary state of two qubits with non-null probability \cite{Bose99,Vidal99}.
However, a very interesting feature emerges when we deal with quantum systems involving more than two qubits: different classes of entanglement arise. For the simplest instance, namely three qubits, there are two inequivalent classes of entanglement \cite{PhysRevA.62.062314}: GHZ states \cite{greenberger1990bell, greenberger1989going} and W states \cite{PhysRevA.62.062314,ZGH1997}, defined as
\begin{equation}
\ket{GHZ}=\frac{1}{\sqrt{2}}\left(\ket{000}+\ket{111}\right), \;\;\;\;\;\;\;\; \ket{W}=\frac{1}{\sqrt{3}} \left(\ket{001}+\ket{010}+\ket{100}\right).
\end{equation}

These two classes are totally inequivalent under Stochastic Local Operations and Communication (hereafter SLOCC). It means that it is impossible to convert any state of a given class into another one and vice-versa. Thus, GHZ and W states constitute Genuinely Entangled states for the case of three qubits.

The idea of GHZ entanglement by itself has its deep origins in the foundations of Quantum Theory. In fact, this family of states was proposed to investigate quantum non-locality beyond Bell's Theorem \cite{greenberger1989going}, as showed in \cite{zukowski1998quest}, the premises of the EPR argument about the incompleteness of Quantum Theory are also inconsistent when applied to GHZ states. As discussed in more details in upcoming sections, the employment of GHZ states led to the possibility to demonstrate the incompatibility between predictions of local realism and quantum mechanics without demanding the usage of an inequality \cite{barnett2009quantum}.

As in the bipartite case, it is also possible to write a GHZ state with arbitrary amount of entanglement. In this case, we start by defining the state
\begin{equation}
\ket{\psi_{000}}=\cos\theta\ket{000}+\sin\theta\ket{111},
\end{equation}
with $\theta=\{0,\pi/2\}$. From local operations on $\ket{\psi_{000}}$, we can construct a GHZ basis.
These states are given by
\begin{equation}
\ket{\psi_{\mu\lambda\omega}}= \sum_{j} (-1)^{\mu j}b_{\mu \oplus j}\ket{j, j \oplus \lambda, j\oplus \omega},
\label{GHZbasis}
\end{equation}
where $b_0=\cos \theta$ e $b_1=\sin \theta$. More explicitly:
\begin{equation}
\ket{\psi_{000}}=\cos\theta\ket{000}+\sin\theta\ket{111},\;\;\;\;\; \ket{\psi_{001}}=\cos\theta\ket{001}+\sin\theta\ket{110},
\end{equation}
\begin{equation}
\ket{\psi_{010}}=\cos\theta\ket{010} + \sin\theta\ket{101},\;\;\;\;\;
\ket{\psi_{011}}=\cos\theta\ket{011} + \sin\theta \ket{100},
\end{equation}
\begin{equation}
\ket{\psi_{100}}=\sin\theta\ket{000} -\cos\theta\ket{111},\;\;\;\;\;
\ket{\psi_{101}}=\sin\theta\ket{001}-\cos\theta\ket{110},
\end{equation}
\begin{equation}
\ket{\psi_{110}}=\sin\theta\ket{010} - \cos\theta\ket{101},\;\;\;\;\;
\ket{\psi_{111}}=\sin\theta\ket{011} - \cos\theta \ket{100}.
\end{equation}

In the same way, we can also define a more general family of entangled W states, given by \cite{schwemmer2015genuine}:
\begin{equation}
\left\vert W_{1}\right\rangle =\sin \theta \cos \varphi \left\vert
001\right\rangle +\sin \theta \sin \varphi \left\vert 010\right\rangle +\cos
\theta \left\vert 100\right\rangle.
\end{equation}

By using local unitary operations, we can generate the other seven states of the W basis:
\begin{equation}
\left\vert W_{2}\right\rangle =+\sin \theta \sin \varphi \left\vert
001\right\rangle -\sin \theta \cos \varphi \left\vert 010\right\rangle +\cos
\theta \left\vert 111\right\rangle,
\end{equation}
\begin{equation}
\left\vert W_{3}\right\rangle =-\sin \theta \sin \varphi \left\vert
100\right\rangle +\cos \theta \left\vert 010\right\rangle +\sin \theta \cos
\varphi \left\vert 111\right\rangle,
\end{equation}
\begin{equation}
\left\vert W_{4}\right\rangle =+\sin \theta \cos \varphi \left\vert
100\right\rangle +\cos \theta \left\vert 001\right\rangle +\sin \theta \sin
\varphi \left\vert 111\right\rangle,
\end{equation}
\begin{equation}
\left\vert W_{5}\right\rangle =+\sin \theta \cos \varphi \left\vert
110\right\rangle +\sin \theta \sin \varphi \left\vert 101\right\rangle +\cos
\theta \left\vert 011\right\rangle,
\end{equation}
\begin{equation}
\left\vert W_{6}\right\rangle =+\sin \theta \sin \varphi \left\vert
110\right\rangle -\sin \theta \cos \varphi \left\vert 101\right\rangle +\cos
\theta \left\vert 000\right\rangle,
\end{equation}
\begin{equation}
\left\vert W_{7}\right\rangle =-\sin \theta \sin \varphi \left\vert
011\right\rangle +\cos \theta \left\vert 101\right\rangle +\sin \theta \cos
\varphi \left\vert 000\right\rangle,
\end{equation}
\begin{equation}
\left\vert W_{8}\right\rangle =+\sin \theta \cos \varphi \left\vert
011\right\rangle +\cos \theta \left\vert 110\right\rangle +\sin \theta \sin
\varphi \left\vert 000\right\rangle.
\end{equation}

Let us consider the partial trace operation on the third qubit in $\psi_{000}$ and $\ket{W_1}$, in order to examine the differences between the two classes. A simple calculation gives us
\begin{equation}
\hat{\rho}_{12}=\tr_3\left(\dyad{\psi_{000}}\right)=\frac{1}{2}\dyad{00}+\frac{1}{2}\dyad{11},
\end{equation}
for the GHZ state. For the W state, we have
\begin{equation}
\hat{\rho}_{12}=\tr_3\left(\dyad{W}\right)=\frac{2}{3}\dyad{\phi_{01}^{(2)}}+\frac{1}{3}\dyad{00}.
\end{equation}
Thus, in the case of a W state, the reduced density operator contains a residual EPR entanglement, in contrast, the same operation on the GHZ state gives a completely disentangled state.

\subsection{Other instances of tripartite entanglement}

In analogy with the bipartite case, Acín and collaborators \cite{Acin2000Schmidt}, defined a Schmidt-like decomposition useful to classify pure three-qubits states \cite{Acin2001p}. These results were after generalized to include mixed states  \cite{acin2001classification}.

In addition to the cases mentioned above, there are several instances of tripartite entanglement. For instance, in \cite{erhard2018experimental}, it is presented a procedure to create a tripartite GHZ entangled states in three levels for every particle. In \cite{Siewert2012}, Siewer and Eltschka use the following generalization of the Werner states
\begin{equation}
\hat{\varrho}_{W}=p\dyad{\psi_{000}}+\frac{1}{8}(1-p)\hat{1}_8,
\end{equation}
to treat the problem of entanglement quantification.

\subsection{Tripartite entanglement in other areas}

It is worth to note that there are several works exploring the idea of entanglement and its classification with other fields in Physics and Mathematics. For instance, there are some works discussing on relationship between entanglement and topology. A connection between Borromean rings and GHZ states was established by Aravind in \cite{aravind1997borromean}. In \cite{bengtsson2016brief}, it is presented a schematic comparison between GHZ and W tripartite states by using knots. There have been efforts to unveil the relation between quantum entanglement and topological entanglement \cite{kauffman2002quantum,asoudeh2004suggestion}. A recent review about this topic can be found in \cite{mironov2019topological}. The idea of entanglement in networks also has been explored.

In \cite{perseguers2010multipartite}, it is presented a strategy for percolation involving GHZ states. In \cite{mccutcheon2016experimental}, it was presented an experimental verification of three and four-party entanglement in quantum networks.

A connection between black-hole physics and quantum entanglement was presented in \cite{duff2013black}. This work also shows that there is a matching between the classification of tripartite entanglement and black holes. A more recent work concerning this topic is \cite{borsten2012black}.

Several works in the literature have been reported focusing on the use of both GHZ and W classes as quantum resources to develop quantum protocols, such as Quantum Teleportation and Superdense Coding. In fact, as we will see, the two classes work in a different way depending on the specific task.

\section{Non locality, Bell's Theorem and GHZ states}

Besides the success of Quantum Theory describing systems in the microscopic world and all the experimental tests in its favour, the foundations of the theory have been widely discussed since its proposal. Albert Einstein, working with Boris Podolsky and Nathan Rosen, for instance, questioned whether Quantum Mechanics could give a complete description of the physical reality. In a paper published in 1935, they established their famous EPR paradox. In their work, EPR conceived an experiment in which Alice and Bob share an ensemble of entangled pairs of qubits, and each of them is able to perform local measurements. Under this scenario, the measurement events are separated by space-like intervals. At each instant Alice may choose one out of two incompatible observables $\hat{A}_1$ or $\hat{A}_2$ (analogously for Bob, $\hat{B}_1$ or $\hat{B}_2$). Assuming that any local action on each particle cannot influence its counterpart (locality), and that measurement results pre-exist for any observable independent of the choice (realism)\footnote{The junction of both premises is known as \textit{local realism} assumption.}, they were able to show that under special cases concerning systems with a high degree of symmetry, two local measurements (one in Alice and the other in Bob's location) allow for the determination of the values associated to the four involved observables, and thus in contradiction with Heisenberg's uncertainty principle. EPR concluded that there is no way in which QM satisfy the local realism assumption, and then there should exist a more general theory possibly described by a set of hidden variables (not available for the experimenter), in analogy to the relation between thermodynamics and statistical mechanics, in which the position of particles in the phase space play the role of hidden variables. Inspired by this, Bell derived a set of conditions (Bell inequalities) satisfied by predictions from any theory based on local hidden variables, which as mentioned before, quantum mechanics violates under certain scenarios \cite{Bell1964}. Since then many efforts have been concentrated to experimentally test Quantum Theory against the local realism hypothesis, with a vast majority in favor of the first one. For a recent revision on the subject, we refer the reader to \cite{Brunner2014}.
\begin{figure}[h!]
\centering
\includegraphics[width=9 cm]{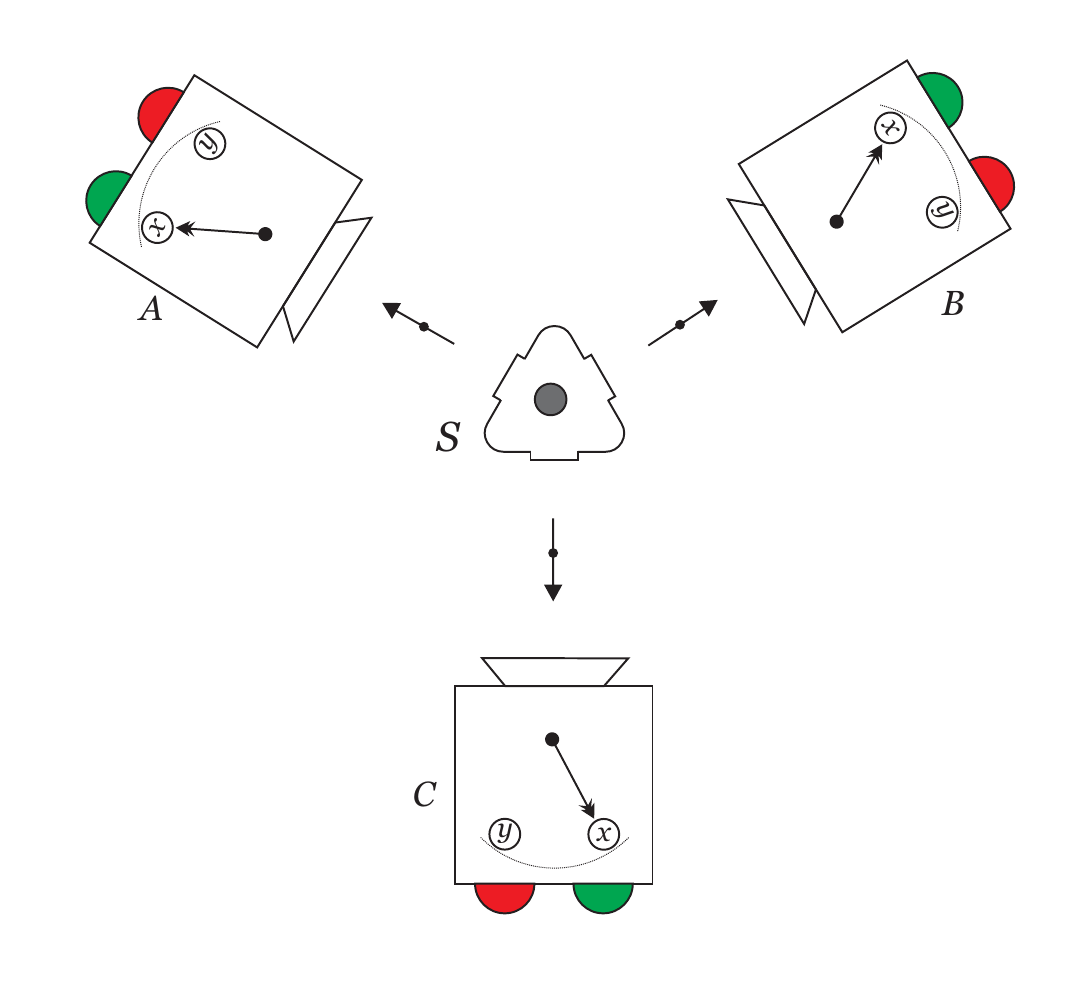}
\caption{ Bell nonlocality in a tripartite scenario. A source S emits three spin $1/2$ particles, traveling to one out of three different detectors located in A, B and C. Each part, namely Alice, Bob and Charlie posses a Stern-Gerlach magnet, and can choose make a $\hat{\sigma}_x$ or $\hat{\sigma}_y$ measurement, obtaining the eigenvalues $+1$ or $-1$, corresponding to turn on green or red lights. }
\end{figure}

Now, let us analyze the conflict between local realism and the predictions of Quantum Mechanics by employing a GHZ state. This analysis was initially proposed by Greenberger, Horne and Zeillinger. Here, we present an alternative version by Mermin \cite{mermin1990quantum}. For an intuitive introduction, see also \cite{barnett2009quantum,lo1998introduction}.
An interesting extension which covers W states is given in \cite{PhysRevA.65.032108}.

First of all, recall the following relations, valid for a single qubit:
\begin{equation}
    \hat{\sigma}_x\ket{0}=\ket{1},\;\;\;\;\;\; \hat{\sigma}_x\ket{1}=\ket{0}, \;\;\;\;\;\;
    \hat{\sigma}_y\ket{0}=i\ket{1},\;\;\;\;\;\; \hat{\sigma}_y\ket{1}=-i\ket{0}.
\end{equation}

Consider three parties, Alice, Bob and Charlie sharing a GHZ state $\ket{\psi_{100}}$.

Let calculate $(\hat{\sigma}_x \otimes \hat{\sigma}_y \otimes \hat{\sigma}_y)\ket{\psi_{100}}$. It means that Alice, Bob and Charlie apply $\hat{\sigma}_x$, $\hat{\sigma}_y$ and $\hat{\sigma}_y$, locally on their qubits respectively. This calculation gives
\begin{equation}
    \hat{\sigma}_x \otimes \hat{\sigma}_y \otimes \hat{\sigma}_y\ket{\psi_{100}}=\frac{1}{\sqrt{2}}\hat{\sigma}_x \otimes \hat{\sigma}_y \otimes \hat{\sigma}_y\left(\ket{000}+\ket{111}\right)=-1\ket{\psi_{100}}.
\end{equation}
Thus the GHZ state is an eigenstate of $\hat{\sigma}_x \otimes \hat{\sigma}_y \otimes \hat{\sigma}_y$, with eigenvalue $-1$. By using the notation employed in \cite{barnett2009quantum}, we can say that the product
\begin{equation}
m_x^A  m_y^B  m_y^C=-1,
\label{m1}
\end{equation}
where, $m_x^A=\pm 1$ indicates the result of the operation $\hat{\sigma}_x$ on Alice's qubit, for example.

We can also calculate $\hat{\sigma}_y \otimes \hat{\sigma}_x \otimes \hat{\sigma}_y\ket{\psi_{100}}$ and $\hat{\sigma}_y \otimes \hat{\sigma}_x \otimes \hat{\sigma}_x\ket{\psi_{100}}$, obtaining the same result, $-1$. In this way we can also write the outcome products as
\begin{equation}
    m_y^A m_x^B m_y^C=-1,~~~~~~~ m_y^A m_y^B m_x^C=-1.
    \label{m3}
\end{equation}
The product of the three terms in Eqs. \ref{m1} and \ref{m3}, holds:
\begin{equation}
     m_x^A  m_y^B  m_y^C m_y^A  m_x^B  m_y^C m_y^A  m_y^B  m_x^C=m_x^A m_x^B m_x^C=-1.
     \label{m_produtos}
\end{equation}
Note that in the last calculations we used the fact that $\left(m_y^A\right)^2=\left(m_y^B\right)^2=\left(m_y^C\right)^2=+1.$
Moreover, the calculation of  $\hat{\sigma}_x\otimes\hat{\sigma}_x\otimes\hat{\sigma}_x\ket{\psi_{100}}$ leads to a quite different result: $+1\ket{\psi_{100}}$, which implies that
\begin{equation}
    m_x^A m_x^B m_x^C=+1.
    \label{m_xxx}
\end{equation}
Thus, we have a contradiction between Eqs. \ref{m_produtos} and \ref{m_xxx}. It indicates the fundamental impossibility to associate pre-determined outcomes for every local measurement performed on a quantum entangled state. In fact, the so called GHZ paradox constitutes the first proof of a possible violation of local realism without the usage of an inequality\footnote{Another interesting proof of the violation local realism without using Bell inequalities is given by the Hardy paradox \cite{Hardy1993}.}.
Furthermore, it is possible to find generalized versions of Bell inequalities for tripartite case. For instance, Mermin, Ardehali, Belinski and Klyshko (MABK) derived independently a set of inequalities capable of testing violation of local realism for a states of $N$ spin-$1/2$ particles \cite{PhysRevLett.65.1838,PhysRevA.46.5375,Belinski_1993}. In addition, Svetlichny made an important contribution to the understanding of genuine tripartite nonlocality \cite{PhysRevD.35.3066}. In his work, it was presented for the first time an inequality that allows the detection of genuine nonlocality in scenarios involving three observers, each capable of performing one out of two dichotomic measurements. In \cite{PhysRevA.66.024102} it is discussed the notion of tripartite entanglement in contrast with that of nonlocality in the context of the Svetlichny inequality. See also \cite{PhysRevA.88.014102}, for a definition of genuine multipartite nonlocality alternative to original Svetlichny's proposal. An more recent discussion about tripartite genuine nonlocality can be found in \cite{paul2016revealing}.

Tripartite entangled states have widely been used to test the previsions of Quantum Theory over Local Hidden Variables Models. For instance, in \cite{pan2000experimental}, was reported a experimental test of quantum nonlocality with GHZ states.
An experimental setup to generate GHZ  states and test the Svetlichny inequality was also reported in \cite{lavoie2009experimental}.
In \cite{erven2014experimental}, it was presented the experimental verification of Mermin's inequality violations by distributing tripartite GHZ states between independent observers. This work closed locality and freedom-of-choice loopholes for three particles.
In \cite{zhang2016experimental}, it was reported an experimental study on quantum nonlocality by dealing with W states.
In \cite{singh2018analysing}, it was made an analysis of nonlocality robustness of GHZ and W states under noisy conditions and weak measurements.
An experimental demonstration of Mermin’s and Svetlichny’s inequalities for GHZ and W states was discussed in \cite{swain2019experimental}.
In \cite{caban2019noise}, it was studied the violation of Svetlichny inequality in the presence of several kinds of noise for the case of GHZ states.

More recently Chaves, Cavalcanti and Aolita, by using the formalism of Bayesian networks have found new different expressions of nonlocality on tripartite states \cite{Chaves2017}.

\section{Quantum information protocols using three-partite entanglement}

Tripartite entanglement can be widely used to execute tasks in the field of quantum information. In this section, we give examples of applicability of this type of entanglement in some protocols.
\begin{figure}[h!]
\centering
\includegraphics[width=7 cm]{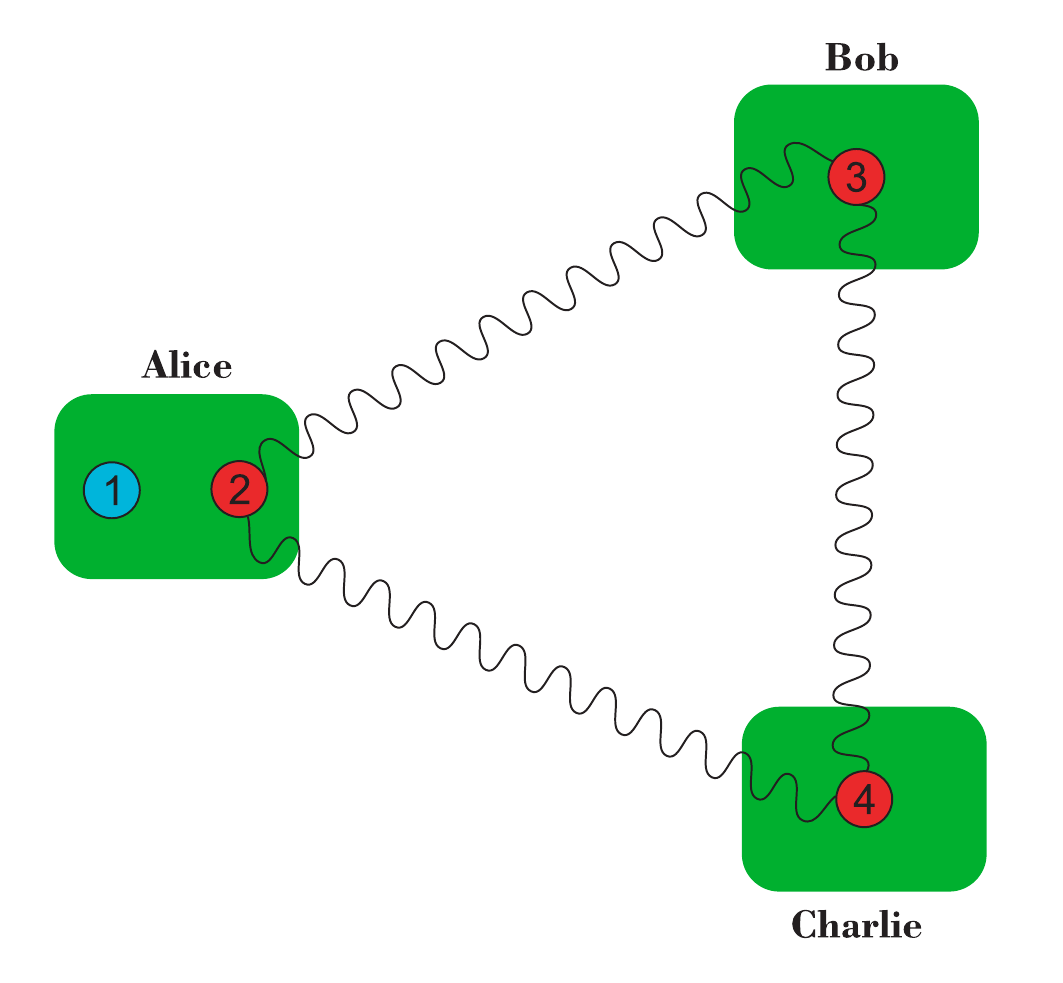}
\caption{A possible scheme to make a teleport of a qubit state by using a three-particle entangled state as the quantum channel.}
\label{canal_tripartite}
\end{figure}

\subsection{Teleportation of a single qubit state by using a GHZ channel and EPR measurements}

\textbf{Attention}: \textit{This section contains a mistake. Please consider the Erratum at the and of the manuscript.}\\

A scheme to make a quantum teleport of a qubit state by using a GHZ state as the channel was reported in \cite{karlsson1998quantum}. In this scheme, there are three users, namely Alice, Bob, and Charlie, as shown in figure \ref{canal_tripartite}. They shared a GHZ state. Alice has a non-shared qubit, whose state she wants to teleport. The state of the system is
\begin{equation}
\ket{\Psi}=[\alpha\ket{0}+\beta\ket{1}]_1 \otimes \frac{1}{\sqrt{2}}[\ket{000}+\ket{111}]_{234}.
\end{equation}
We can write this state in a more compact form as
\begin{equation}
\ket{\Psi}=\sum_{i=0}^1 \alpha_i \ket{i}_1 \otimes \sum_{j=0}^1 \frac{1}{\sqrt{2}} \ket{jjj}_{234}.    \end{equation}
In this case, Alice makes a Bell-measurement on the qubits 1 and 2.
This way, let us separate the qubits in the following way:
\begin{equation}
\ket{\Psi}=\frac{1}{\sqrt{2}}\sum_{i,j} \alpha_i \ket{ij}_{12} \ket{jj}_{34}.
\end{equation}

Let us calculate the projection of $\ket{\Psi}$ into the $m,n$ element of the EPR basis on qubits 1 and 2, $\bra{\phi_{mn}^{(2)}}_{12}$. It holds
\begin{equation}
     \bra{\phi_{mn}}_{12}\hspace{0.05cm}\ket{\Psi}=\ket{\eta}_{34}=\frac{1}{2}\sum_{ijk}\alpha_i (-1)^{mk}\left\langle k,k \oplus n \vert i,j \right\rangle_{12}\otimes \ket{j,j}_{34}.
\end{equation}
 As result, Charlie and Bob, now, share an EPR state:
\begin{equation}
    \ket{\eta}_{34}=\ket{\eta_{mn}}_{34}=\sum_{k=0}^1 (-1)^{mk}\alpha_k \ket{k, k \oplus n}_{34}.
\end{equation}
This EPR state it is not necessarily $\ket{\Phi^+}$. Actually, it can be any one of the four sttates of the EPR basis. Which state Charlie and Bob will share have a dependence on Alice's measurement result. It is important to note that the new state shared by Charlie and Bob it is not a perfect EPR state because the coefficients of this state are $\alpha_0$ and $\alpha_1$.
If Alice obtains the state $\ket{\psi_{00}},$ for instance, Bob and Charlie share the state
\begin{equation}
\ket{\eta_{00}}_{34}=\sum_k \alpha_k \ket{k,k}=\alpha_0 \ket{00}+\alpha_1 \ket{11}.
\label{eta00_eq}
\end{equation}
\begin{table}
\caption{Results after Alice's measurement and the states that Bob and Charlie shares.}
\centering
\begin{tabular}{ccc}
\toprule Alice's result & & State shared by Bob and Charlie\\
\midrule $m$ & $n$ & state \\
\midrule 0 & 0 & $\ket{\eta_{00}}= \alpha_0\ket{00}+ \alpha_1\ket{11} $ \\
\midrule 0 & 1 & $\ket{\eta_{01}}=\alpha_0\ket{01}+ \alpha_1\ket{10} $ \\
\midrule 1 & 0 & $\ket{\eta_{10}}=\alpha_0\ket{00}- \alpha_1\ket{11} $ \\
\midrule 1 & 1 & $\ket{\eta_{11}}=\alpha_0\ket{01}-\alpha_1 \ket{10} $ \\
\bottomrule
\end{tabular}
\label{eta}
\end{table}
The table \ref{eta} exhibits the states that can be shared by Bob and Charlie after Alice's measurement. Let us explore in more details the case $\ket{\eta_{00}}.$ As shown in \cite{karlsson1998quantum}, to proceed with the protocol, Bob or Charlie should make a measurement on a single qubit. Let us suppose that Alice wants to teleport the state of her qubit to Charlie. Then, she asks Bob to make a measurement on his qubit. Bob uses the following basis:
\begin{equation}
\ket{0}_3=\sin\theta \ket{x_0}_3+\cos\theta \ket{x_1}_3, \;\;\;\;\;
\ket{1}_3=\cos\theta \ket{x_0}_3-\sin\theta \ket{x_1}_3.
\end{equation}
Replacing these expressions on the equation \ref{eta00_eq}, we have:
\begin{equation}
\ket{\eta_{00}}_{34}=\alpha_0 (\sin\theta\ket{x_1}_3+\cos\theta\ket{x_2}_3)\ket{0}_4 +\alpha_1 (\cos\theta\ket{x_1}_3-\sin\theta\ket{x_2}_3)\ket{0}_4.
\end{equation}
We can reorganize the expression as
\begin{equation}
\ket{\eta_{00}}_{34}=\ket{x_1}_3(\alpha_0 \sin\theta\ket{0}_4+\alpha_1\cos\theta\ket{1}_4)+\ket{x_2}_3(\alpha_0\cos\theta \ket{0}_4-\alpha_1\sin\theta \ket{1}_4).
\end{equation}
Thus, if Bob measures $\ket{x_1},$ the state (up to normalization) is $\alpha_0 \sin\theta\ket{0}_4+\alpha_1\cos\theta\ket{1}_4$. The table \ref{eta} shows all possibilities after Alice's measurement and Bob's measurement.
\begin{table}[]
\caption{Alice's and Bob's results; and the final state obtained by Charlie.}
\centering
\begin{tabular}{cccc}
\toprule Alice's result & & Bob's result   & State (unnormalized) of  Charlie's qubit\\
\midrule $m$ & $n$ & $\ket{x_i}$  & state \\
\midrule 0 & 0 & $\ket{x_0}$  & $ \alpha_0\sin\theta\ket{0}+ \alpha_1\cos\theta\ket{1} $ \\
\midrule 0 & 0 & $\ket{x_1}$ & $\alpha_0\cos\theta\ket{0}- \alpha_1\sin\theta\ket{1} $ \\
\midrule 0 & 1 & $\ket{x_0}$ & $\alpha_0\sin\theta\ket{1}+ \alpha_1\cos\theta\ket{0} $ \\
\midrule 0 & 1 & $\ket{x_1}$ & $\alpha_0\cos\theta\ket{1}-\alpha_1\sin\theta \ket{0} $ \\
\midrule 1 & 0 & $\ket{x_0}$ & $\alpha_0\sin\theta\ket{0}-\alpha_1\cos\theta \ket{1} $ \\
\midrule 1 & 0 & $\ket{x_1}$ & $\alpha_0\cos\theta\ket{0}+\alpha_1\sin\theta \ket{1} $ \\
\midrule 1 & 1 & $\ket{x_0}$ & $\alpha_0\sin\theta\ket{1}-\alpha_1\cos\theta \ket{0} $ \\
\midrule 1 & 1 & $\ket{x_1}$ & $\alpha_0\cos\theta\ket{1}+\alpha_1\sin\theta \ket{0} $ \\
\bottomrule
\end{tabular}
\label{mnx}
\end{table}
\begin{table}
\caption{Alice's and Bob's results; and the corresponding operations needed to recover the desired state.}
\centering
\begin{tabular}{cc}
\toprule  Result: \hspace{0.05cm} $(m,n,i)$ & Operation \\
\midrule $(0,0,0)$ & $\hat{I}$  \\
\midrule $(0,0,1)$ & $\hat{\sigma}_z$  \\
\midrule $(0,1,0)$ & $\hat{\sigma}_x$  \\
\midrule $(0,1,1)$ & $\hat{\sigma}_z\hat{\sigma}_x$  \\
\midrule $(1,0,0)$ & $\hat{\sigma}_z$ \\
\midrule $(1,0,1)$ & $\hat{I}$ \\
\midrule $(1,1,0)$ & $\hat{\sigma}_z\hat{\sigma}_x$  \\
\midrule $(1,1,1)$ & $\hat{\sigma}_x$ \\
\bottomrule
\end{tabular}
\label{unitarias1}
\end{table}
We can write the state of Bob's qubit, before his measurement, as
\begin{equation}
\ket{k}_3=\sum_{j=0}^1(-1)^{jk}b_{j \oplus k}\ket{x_j}_3.
\end{equation}
Here, we have defined
\begin{equation}
b_0 \equiv \sin\theta, \hspace{0.4cm} b_1=\cos\theta.
\end{equation}
This way,
\begin{equation}
\ket{\eta_{mn}}_{34}=\sum_{k=0}^1 (-1)^{mk}\alpha_k \left(\sum_{j=0}^1(-1)^{jk}b_{j \oplus k}\ket{x_j}_3\right)\ket{k \oplus n}_4.
\end{equation}
Essentially, this expression contains the same information showed on the table \ref{eta}.
To obtain the final state of Charlie's qubit, we can calculate the projection of the above state on $\ket{x_{\ell}}$, where $\ell=0,1.$
\begin{equation}
 \ket{\chi_{mn0}}_4=\bra{x_0}_3 \hspace{0.05cm}\ket{\eta_{mn}}_{34}=\sum_{k=0}^1 (-1)^{mk}\alpha_k b_k\ket{k\oplus n}.
\end{equation}
\begin{equation}
 \ket{\chi_{mn1}}_4 = \bra{x_1}_3 \hspace{0.05cm}\ket{\eta_{mn}}_{34}=\sum_{k=0}^1 (-1)^{(m\oplus 1)k}\alpha_k b_{k\oplus 1}\ket{k\oplus n}.
\end{equation}
The indexes $mn0$ and $mn1$ are related with the measurements of Alice and Bob on their qubits. These index are the same showed in table \ref{mnx}. To recover the desired state, Charlie needs to apply a unitary transformation on his qubit. Which unitary he will use, depends of $(m,n,i)$. The table \ref{unitarias1} shows  list of the operations for all results. For the case $(0,0,0)$, for example, the state (up to normalization) is
\begin{equation}
\ket{\chi_{000}}=\alpha_0 \sin\theta\ket{0}_4+\alpha_1\cos\theta \ket{1}_4.
\end{equation}
In this case, the operation is the identity. The normalized state is
\begin{equation}
\ket{\chi_{000}}_f = \frac{\alpha_0 \sin\theta\ket{0}_4+\alpha_1\cos\theta \ket{1}_4}{\sqrt{|\alpha_0|^2\sin^2\theta +|\alpha_1|^2\cos^2\theta}}.
\end{equation}
Let's see what happens when the result is $(0,0,1).$ In this case, the state, up to normalization, is
\begin{equation}
\ket{\chi_{001}}=\alpha_0 \cos\theta\ket{0}_4-\alpha_1\sin\theta \ket{1}_4;
\end{equation}
Now, Charlie needs to apply the $\hat{\sigma}_z$ operation on his qubit to recover the desired state. We can note that
\begin{equation}
    \hat{\sigma}_z \ket{\chi_{001}}=\alpha_0 \cos\theta\ket{0}_4+\alpha_1\sin\theta \ket{1}_4.
\end{equation}
After the normalization, we have
\begin{equation}
\ket{\chi_{001}}_f = \frac{\alpha_0 \cos\theta\ket{0}_4+\alpha_1\sin\theta \ket{1}_4}{\sqrt{|\alpha_0|^2\cos^2\theta +|\alpha_1|^2\sin^2\theta}}.
\end{equation}
All the results can be viewed on tables \ref{eta} and \ref{unitarias1}.
Another work dealing with GHZ states as the quantum channel can be accessed in \cite{almeida1998one}.

\subsection{Teleportation of a single qubit state: GHZ channel and measurement}

Another possibility to use GHZ states in quantum teleportation is by employing a GHZ channel and making a GHZ measurement on Alice's side.
An example of a protocol dealing with GHZ measurements can be viewed in \cite{choudhury2017simultaneous}.
Let us suppose that, now, Alice wants to teleport of a single qubit, and shares a GHZ entangled state with Bob. Let us consider that Alice has two qubits of the shared GHZ state, and Bob has another one. Thus, now Alice has three qubits and Bob has one. To diversify the mathematical approach used here, let us now write the states by using the formalism of the density matrix. This problem has been discussed in \cite{moreno2018using}, where was considered a maximally entangled state as the channel as well in the measurement. It was showed that this procedure can help to improve the quality of the protocol on the occurrence of bit-flip error. Here, let us consider non-maximally entangled GHZ states. It is interesting to consider such states, because, we can see how the quality of the entanglement affects the final results. The effect of dealing with non-maximally GHZ entangled states also can be viewed in other scenarios. In \cite{man2007quantum}, for instance, has addressed the problem of sharing a multiqubit state using non-maximally GHZ entangled states.
\begin{figure}[h!]
\centering
\includegraphics[width=7 cm]{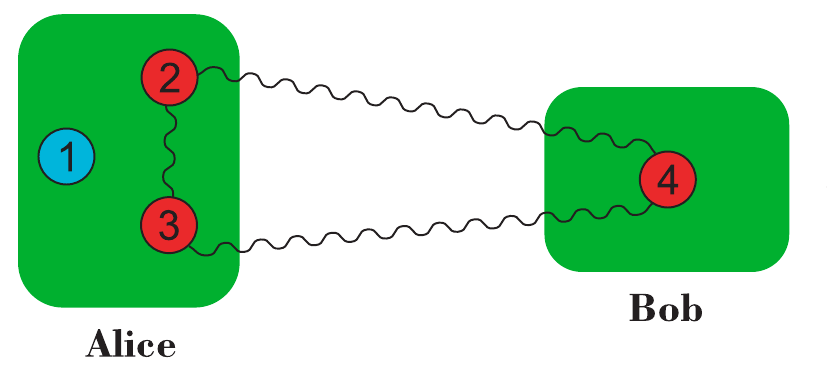}
\caption{Scheme to teleport a single qubit state by using a GHZ state as the quantum channel. A GHZ measurement it is taken on the qubits 1, 2, 3.}
\label{canal_medida_GHZ}
\end{figure}
Alice wants to teleport the state
\begin{equation}
 \ket{\phi}=c_0\ket{0}+c_1\ket{1}=\sum_{i=0}^1 c_i \ket{i}.
\end{equation}
 The corresponding density operator is given by
\begin{equation}
\hat{\rho}_{1}= \sum_{ij}c_i c_j^{*}\ket{i}\bra{j}.
\end{equation}
Let us consider that the channel is given by $\ket{\Phi^+}_{ghz}=\beta_0 \ket{000}+\beta_1 \ket{111},$ where $\beta_0=\cos \theta$ and $\beta_1=\sin \theta.$ We can write this state as
\begin{equation}
\ket{\Phi^+}_{ghz}=\sum_{k=0}^1 \beta_k\ket{kkk}.
\end{equation}
The corresponding density operator is
\begin{equation}
\hat{\rho}_{234}=\sum_{k \ell}\beta_k \beta_\ell \ket{kkk}\bra{\ell \ell \ell}.
\end{equation}
The state of the whole system is
\begin{equation}
\hat{\rho}=\hat{\rho}_1 \otimes \hat{\rho}_{234}=\sum_{ijk\ell}c_i c_j^* \beta_k \beta_{\ell}\ket{ikkk}_{1234}\bra{j\ell \ell \ell}_{1234}.
\end{equation}
Let us separate the qubits that will taken on the GHZ measurement.
\begin{equation}
\hat{\rho}=\hat{\rho}_1 \otimes \hat{\rho}_{234}=\sum_{ijk\ell}c_i c_j^* \beta_k \beta_{\ell}\ket{ikk}_{123}\bra{j\ell \ell}_{123}\otimes \ket{k}_4\bra{\ell}_4.
\end{equation}
To the measurement, we will consider the basis introduced in Eq. \ref{GHZbasis}.

After the measurement, the unnormalized state is given by
\begin{equation}
    \tilde{\rho}_I(\mu,\lambda,\omega)=\sum_{j'k'}(-1)^{\mu j'}(-1)^{\mu k'}b_{\mu \oplus j'}b_{\mu \oplus k'}\beta_{k' \oplus \lambda}\beta_{j' \oplus \lambda}\delta_{\lambda,\omega}c_{k'}c_{j'}^*\ket{k' \oplus \lambda}_4\bra{j' \oplus \lambda}_4.
\end{equation}
This state depends on the parameters related to the measurement, $(\mu,\lambda,\omega)$.
For the case where $(\mu,\lambda,\omega)=(0,0,0)$, for example, we have
\begin{equation}
    \tilde{\rho}_I(0,0,0)=\sum_{j'k'}(b_{j'}b_{k'}\beta_{j'}\beta_{k'}\delta_{\lambda,\omega})c_{k'}c_{j'}^*\ket{k'}\bra{j'}.
\end{equation}
More explicitly,
\begin{equation}
    \tilde{\rho}_I(0,0,0)= b_0^2\beta_0^2|c_0|^2 \ket{0}\bra{0}+b_1^2\beta_1^2|c_1|^2 \ket{1}\bra{1}+ b_0b_1\beta_0\beta_1(c_0c_1^*\ket{0}\bra{1} + c_1c_0^*\ket{1}\bra{0}).
\end{equation}
By the other hand,
\begin{equation}
    \tilde{\rho}_I(1,1,1)=\sum_{j'k'}(-1)^{ j'}(-1)^{ k'}b_{j' \oplus 1}b_{k' \oplus 1}\beta_{k' \oplus 1}\beta_{j' \oplus 1}\delta_{\lambda,\omega}c_{k'}c_{j'}^*\ket{k' \oplus 1}_4\bra{j' \oplus 1}_4,
\end{equation}
or
\begin{equation}
    \tilde{\rho}_I(1,1,1)= b_1^2\beta_1^2|c_0|^2 \ket{1}\bra{1}+b_0^2\beta_0^2|c_1|^2 \ket{0}\bra{0}- b_0b_1\beta_0\beta_1(c_1c_0^*\ket{0}\bra{1} + c_0c_1^*\ket{1}\bra{0}).
\end{equation}
The normalized state after the measurement is given by
\begin{equation}
\hat{\rho}_I= \hat{\rho}_I(\mu,\lambda,\omega)= \frac{\hat{P}\hat{\rho}_0 \hat{P}^{\dagger}}{P}=\frac{ \tilde{\rho}_I}{P},
\end{equation}
where
\begin{equation}
\hat{P}=\hat{P}_{\mu \lambda \omega}= (\ket{\Phi^{\mu}_{\lambda}}\bra{\Phi^{\mu}_{\lambda \omega}})_{123}
\end{equation}
is the projector associated to GHZ measurement, and
\begin{equation}
P=P_{\mu \lambda \delta \omega}= Tr(\hat{P}\hat{\rho}_0)
\end{equation}
corresponds to probability of each outcome.

Each outcome $(\mu,\lambda,\omega)$ requires a unitary transformation, in order to recover the desired state.  In the case $(0,0,0)$, the operation it is just the identity. In the case $(1,1,1)$, Bob needs to apply $\hat{\sigma}_z\hat{\sigma}_x.$
Looking to the expression to $\tilde{\rho}_I$, it is possible to suggest that the general form of  operation needed is given by
\begin{equation}
    \hat{U}=\hat{U}_{\mu \lambda \omega}= (\hat{\sigma}_z)^{\mu}(\hat{\sigma}_x)^{\lambda}.
\end{equation}
After the measurement, by using a classical channel, Alice informs Bob which $\mu$ and $\lambda$ are needed to recover the desired state.
We can check if this operation works by calculating $\hat{U}\tilde{\rho}_I \hat{U}^{\dagger}.$
\begin{equation}
    \tilde{\rho}_f=\hat{U}\tilde{\rho}_I \hat{U}^{\dagger}=\sum_{j'k'}b_{\mu \oplus j'}b_{\mu \oplus k'}\beta_{k' \oplus \lambda}\beta_{j' \oplus \lambda}\delta_{\lambda,\omega}c_{k'}c_{j'}^*\ket{k'}\bra{j'}.
\end{equation}
This way, the unitary corrects the factor $k' \oplus \lambda$ on the \textit{ket} and the factor $(-1)^{\mu k'}$. However, it is impossible to correct the factors involving $b$ and $\beta$.

The fidelity correspondent to a specific outcome is
\begin{equation}
F_{\mu \lambda \omega}= Tr(\hat{\rho}_{in}\tilde{\rho}_f).
\end{equation}
The average fidelity is
\begin{equation}
\overline{F}=\sum_{\mu \lambda \omega}P_{\mu \lambda \omega}F_{\mu \lambda \omega}=\sum_{\mu \lambda \omega}P_{\mu \lambda \omega}Tr[\hat{\rho}_{in}\hat{\rho}_f].
\end{equation}
Here, ${\rho}_f$ is the final state, after the unitary transformation.
When we deal with maximally entangled states, we know that the probability of obtaining a specific outcome it is the same than the others (the probability it is the same for all outcomes.) It is not true in more general cases. Then, this average fidelity takes into account that, when we deal with non-maximally entangled states, the probability of obtaining an outcome depends on the amount of entanglement. Also, each outcome has a different probability. In this case, we need to sum all the contributions of each outcome times the correspondent statistical weight $P_{\mu \lambda \omega}$.
But,
\begin{equation}
\hat{\rho}_f = \frac{\hat{U}\tilde{\hat{\rho}}_I \hat{U}^{\dagger}}{P_{\mu \lambda \omega}}= \frac{\tilde{\rho}_f}{P_{\mu \lambda \omega}}.
\end{equation}
Thus,
\begin{equation}
\overline{F}= \sum_{\mu \lambda \omega}Tr[\hat{\rho}_{in}\tilde{\rho}_f],
\end{equation}
or
\begin{equation}
\overline{F}=\sum_{k \ell \mu \nu  \lambda \omega}|c_k|^2 |c_{\ell}|^2  b_{ \mu \oplus k}^* b_{\mu \oplus \ell}\beta_{k \oplus \lambda}\beta_{\ell \oplus \lambda}.
\end{equation}
Making all the sums, we have
\begin{equation}
\overline{F}=|c_0|^4+|c_1|^4+2|c_0|^2|c_1|^2 \left(b_0b_1\beta_0\beta_1\right) .
\end{equation}
The expression above depends on the initial state. In order to obtain a quantity independent of the parameters of the initial state, let us consider that the initial state can be parametrized as
\begin{equation}
\ket{\Psi}_{in}= \left|c_0\right| \ket{000} + \left|c_1\right|e^{i\varphi}\ket{111},
\end{equation}
in a such way we can now calculate a average on all possible input states:
\begin{equation}
\langle \overline{F} \rangle = \frac{1}{2\pi} \int_0^{2\pi}\int_0^1 \overline{F}(|c_0|^2, \varphi) d|c_0|^2 d\varphi.
\end{equation}
Evaluating the integral, we obtain
\begin{equation}
\langle \overline{F} \rangle = \frac{2}{3}+ \frac{1}{3}\sin(2\theta) \sin(2\phi).
\end{equation}
In this expression, the first term corresponds to the classical contribution, and the second one depends on the entanglement of the channel and of measurement basis. Essentially, it has the same form that the fidelity for the standard teleportation protocol, using EPR channel and measurement. However, as pointed in \cite{moreno2018using}, in the presence of bit-flip noise, GHZ states allow correction of error. This way, it is possible to improve fidelity.
\begin{figure}[h!]
\centering
\includegraphics[width=8.5 cm]{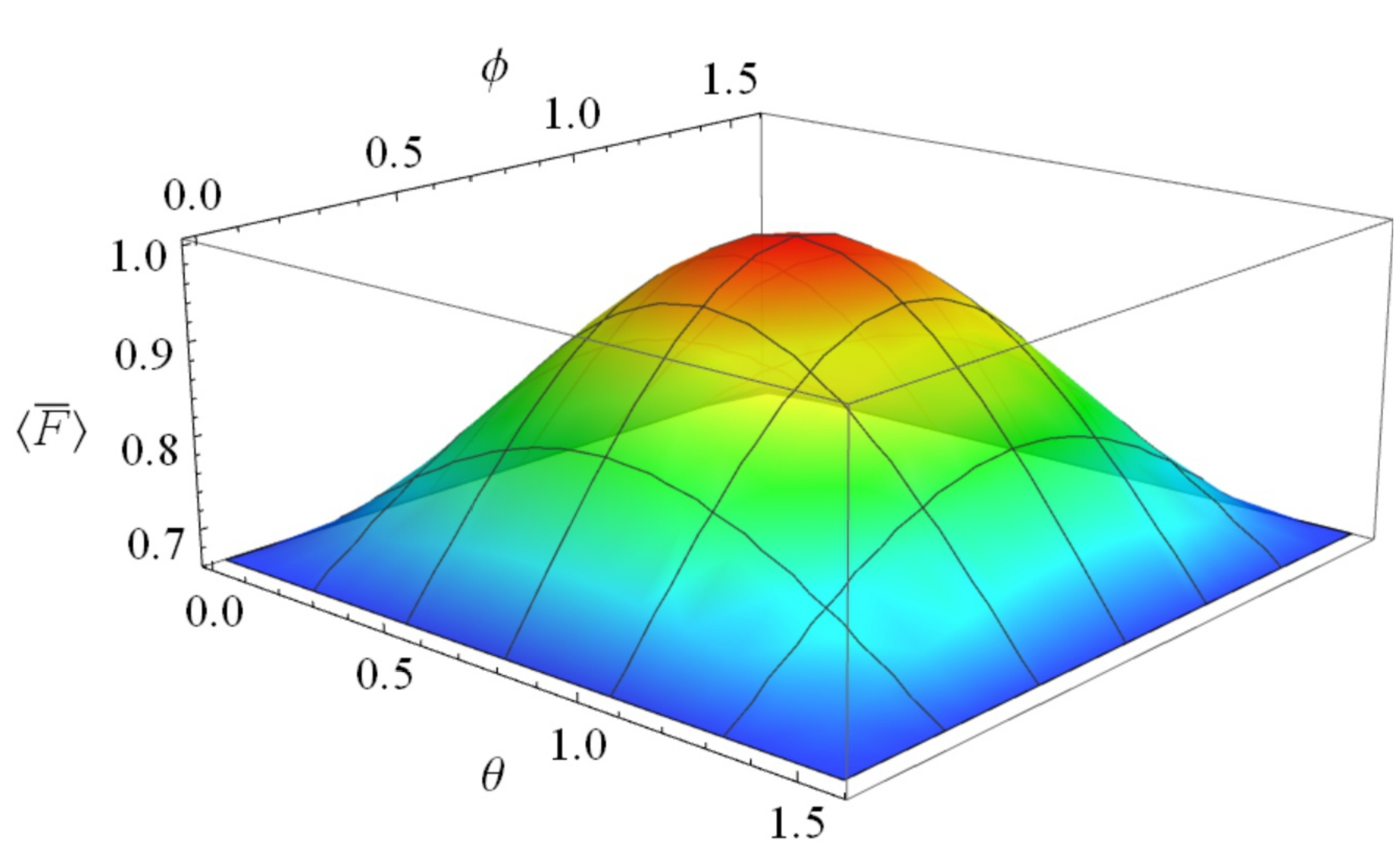}
\caption{Graph of the quantum fidelity corresponding to the case of a single-qubit state teleport by using a GHZ channel, as function of entanglement of the channel as well the entanglement of measurement basis.}
\label{fidelidade_canal_medida_GHZ}
\end{figure}
The figure \ref{fidelidade_canal_medida_GHZ} shows the plot of the fidelity as a function of $\theta$ and $\phi$. When the states on the channel and in the measurement basis are maximally entanglement, the maximum of fidelity is reached, corresponding to $\langle \overline{F} \rangle=1$, the perfect case.

\subsection{Teleportation of a two-qubit state}

An interesting quantum teleportation protocols consist in making a teleport of a two-qubits entangled state \cite{wang2019probabilistic}. In \cite{gorbachev2000quantum}, was proposed a scheme to teleport an EPR-like state by using GHZ states as the channel. Others schemes also have been proposed, like in \cite{shi2000probabilistic, tsai2010teleportation}. Recently, it was showed a scheme to teleport an arbitrary two-qubit state by using two GHZ states \cite{li2019quantum}.
In \cite{dai2004probabilistic}, it was exhibited a protocol to teleport a two-qubit state by using a both GHZ and W states simultaneously as the quantum channel.
The teleport of an EPR state by a non-maximally entangled GHZ quantum channel was discussed in \cite{wang2018teleportation}.
In \cite{hassanpour2016bidirectional}, was shown a scheme to make bidirectional teleportation of an EPR state using GHZ states. In \cite{zou2017multihop}, a protocol dealing with a composite channel using EPR and GHZ states is presented, to execute a multihop teleport of a two-qubit state. In the reference \cite{wang2018probabilistic}, it was presented a scheme to teleport an arbitrary two-qubit state by using a channel consisting of a tripartite entangled state (GHZ or W) and an EPR state.
\begin{figure}[h!]
\centering
\includegraphics[width=7 cm]{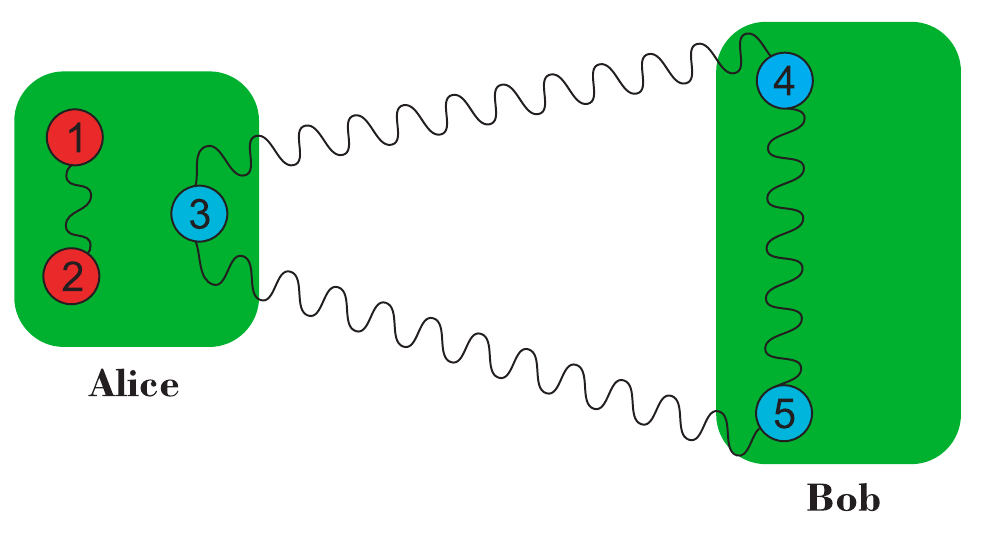}
\caption{An EPR state can be teleported by using a GHZ state as the quantum channel. A measurement on GHZ basis also it is considered.}
\label{teleport_EPR}
\end{figure}
The figure \ref{teleport_EPR} shows a possible scheme to teleport a EPR state. A GHZ state can be used as the quantum channel, and a measurement in a GHZ-basis on the qubits 1, 2 and 3 it is taken.
The state of the system is
\begin{equation}
    (a_0\ket{00}+a_1\ket{11})_{12}\otimes \frac{1}{\sqrt{2}}(\ket{000}+\ket{111})_{345}.
\end{equation}

\subsection{Teleportation of a GHZ state}

\begin{figure}[h!]
\centering
\includegraphics[width=11.0 cm]{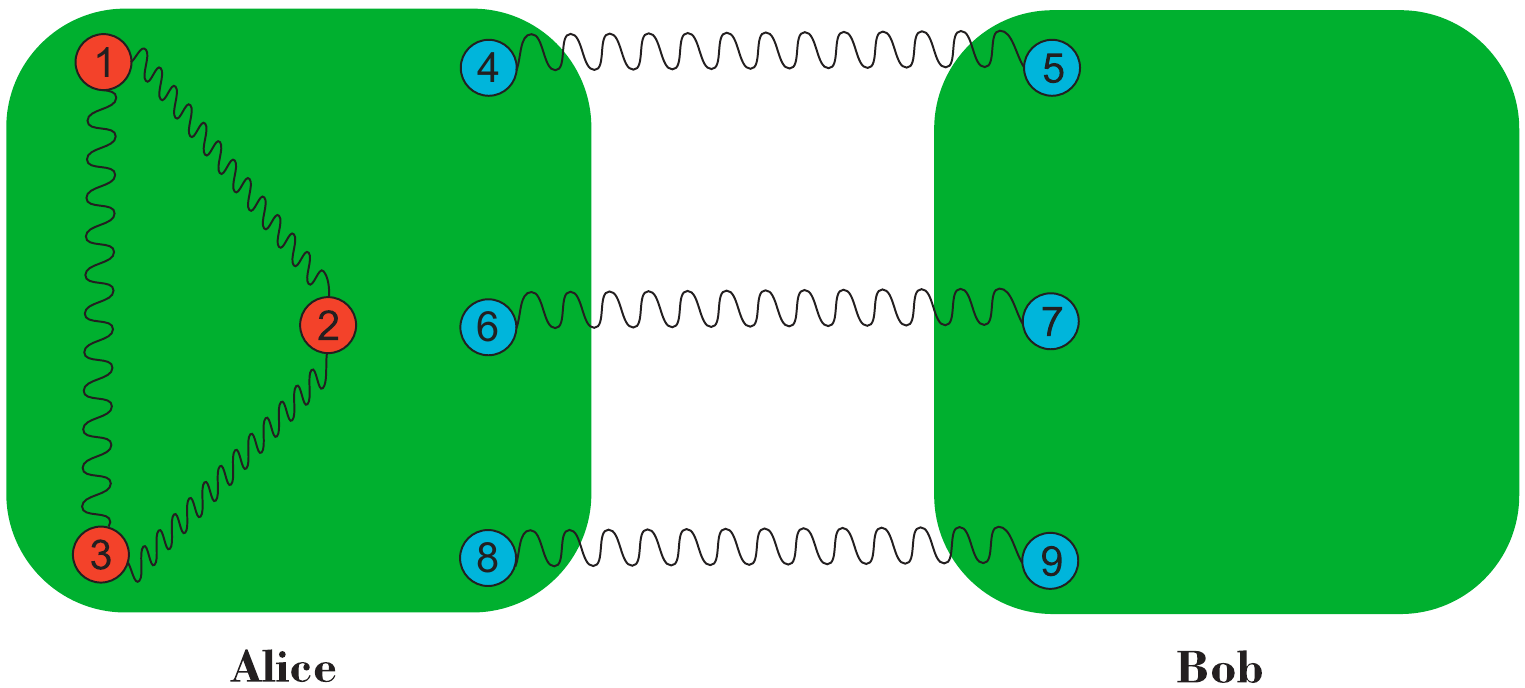}
\caption{A possible scheme to teleport a GHZ state. Three EPR states are used as the quantum channel, and EPR measurements are taken on qubits 1, 4 and 2, 6 and 3, 8. At the end, the qubits 5, 7, 9 are in a GHZ state.}
\label{esquemaGHZ_canaisEPR}
\end{figure}

Another possibility of quantum teleportation protocol it is the teleport of a GHZ state. It is an interesting issue, because besides make a teleport of information itself, the protocol also provides quantum entanglement to other locations. It is relevant in the sense that the distribution of entanglement could be used, in principle, to connect several distant users in a quantum network, for example. Several works have been addressed to this problem. In \cite{xia2010teleportation}, was investigated the teleportation of GHZ state with N photons by using a two-photon entangled state as the quantum channel. A scheme to teleport a GHZ state via entanglement-swapping is showed in \cite{hong2001probabilistic}. The teleportation of a GHZ state by using two W entangled states as the quantum channel was considered in \cite{xiu2006probabilistic}.
GHZ states also can be transmitted through a multihop teleportation scheme by using Bell’s states as intermediate quantum channels \cite{choudhury2018teleportation}.
In \cite{jin2002quantum}, was proposed a scheme to teleport a GHZ-like state by employing three pairs of non-maximally entangled states as the quantum channel. The configuration to make the protocol is illustrated in figure \ref{esquemaGHZ_canaisEPR}. In \cite{fang2003probabilistic}, a similar protocol to the last one was presented, but this time to teleport an arbitrary tripartite state.
The figure \ref{esquemaGHZ_canaisEPR} shows a possible scheme to make a teleport of a GHZ state by using three EPR states as the quantum channel. The state of the system is given by
\begin{equation}
    \ket{\Psi}=\ket{\phi}_{123}\otimes \ket{\eta}_{45} \otimes \ket{\chi}_{67}\otimes \ket{\zeta}_{89},
\end{equation}
where
\begin{equation}
\ket{\phi}_{123}=a_0\ket{000}_{123}+a_1\ket{111}_{123},
\end{equation}
\begin{equation}
\ket{\eta}_{45}=\sum_j b_j\ket{jj}_{45}, \hspace{0.5cm}  \ket{\chi}_{67}=\sum_k c_k\ket{kk}_{67}, \hspace{0.5cm} \ket{\zeta}_{89}=\sum_{\ell} x_{\ell}\ket{\ell \ell}_{89}.
\end{equation}
In this scheme, Alice makes three EPR measurements on her side. The measurements are taken on the particles (1,4), (2,6) and (3,8). The reference \cite{jin2002quantum} shows the procedure in details to finish the protocol.

\subsection{Teleportation of a single qubit state using a W channel}

In \cite{shi2002teleportation}, was reported a possible scheme to deal with a single qubit state teleportation by employing a W state as the channel and an EPR measurement. In this work, was showed that W states are suitable to make the protocol probabilistically. In this work, there were some wrong calculations about the probabilities. It was set by \cite{joo2002comment} and \cite{shi2002reply}. However, besides the conclusion presented in \cite{shi2002teleportation} is correct, the calculations for the case of a W state with the generic state it is not quite precise. In the following, we explicit the correct calculations for a teleport of a qubit state by using a generic W state. The state to be teleported is
\begin{equation}
\ket{\phi}_1=\alpha\ket{0}_1+\beta\ket{1}_1.
\end{equation}
The channel is
\begin{equation}
\ket{W}_{234}=a\ket{100}_{234}+b\ket{010}_{234}+c\ket{001}_{234}.
\end{equation}
The general state of the system is
\begin{equation}
\ket{\Psi}=[\alpha\ket{0}+\beta\ket{1}]_1 \otimes \left[a\ket{100}+b\ket{010}+c\ket{001}\right]_{234}.
\end{equation}
More explicitly, we have
\begin{align}
\ket{\Psi}&= \alpha a\ket{0100}_{1234}+\alpha b \ket{0010}_{1234}+\alpha c \ket{0001}_{1234} \notag \\
&+ \beta a\ket{1100}_{1234}+\beta b \ket{1010}_{1234}+\beta c \ket{1001}_{1234}.
\end{align}
Now, we can write the states of qubits 1 and 2 in terms of the EPR basis.
\begin{align}
 \left\vert \Psi \right\rangle &=\frac{\alpha a }{\sqrt{2}}%
\left( \left\vert \psi ^{+}\right\rangle +\left\vert \psi ^{-}\right\rangle
\right) _{12}\left\vert 00\right\rangle _{34}+\frac{\alpha b }{\sqrt{2}}\left( \left\vert \Phi ^{+}\right\rangle +\left\vert \Phi
^{-}\right\rangle \right) _{12}\left\vert 10\right\rangle _{34}  \notag \\
& +\,\frac{\alpha c }{\sqrt{2}}\left( \left\vert \Phi ^{+}\right\rangle
+\left\vert \Phi ^{-}\right\rangle \right) _{12}\left\vert 01\right\rangle
_{34}+\frac{\beta a }{\sqrt{2}}\left( \left\vert \Phi
^{+}\right\rangle -\left\vert \Phi ^{-}\right\rangle \right) _{12}\left\vert
00\right\rangle _{34}  \notag \\
& +\frac{\beta b }{\sqrt{2}}\left( \left\vert \psi
^{+}\right\rangle -\left\vert \psi ^{-}\right\rangle \right) _{12}\left\vert
10\right\rangle _{34}+\frac{\beta c }{\sqrt{2}}\left( \left\vert \psi
^{+}\right\rangle -\left\vert \psi ^{-}\right\rangle \right) _{12}\left\vert
01\right\rangle _{34};
\end{align}
We can write this expression in the following way:
\begin{align}
 \left\vert \Psi \right\rangle&=\frac{\left\vert \Phi ^{+}\right\rangle_{12}}{\sqrt{2}}[\alpha b \left\vert 10\right\rangle + \alpha c\left\vert 01\right\rangle+\beta a \left\vert 00\right\rangle]_{34} +\frac{\left\vert \Phi ^{-}\right\rangle_{12}}{\sqrt{2}}[\alpha b \left\vert 10\right\rangle + \alpha c\left\vert 01\right\rangle-\beta a \left\vert 00\right\rangle]_{34} \notag \\
& +\frac{\left\vert \psi ^{+}\right\rangle_{12}}{\sqrt{2}}[\alpha a \left\vert 00\right\rangle + \beta b\left\vert 10\right\rangle+\beta c \left\vert 01\right\rangle]_{34}  +\frac{\left\vert \psi ^{-}\right\rangle_{12}}{\sqrt{2}}[\alpha a \left\vert 00\right\rangle - \beta b\left\vert 10\right\rangle-\beta c \left\vert 01\right\rangle]_{34}. \notag \\
\end{align}
By separating the qubits 3 and 4, results
\begin{align}
 \left\vert \Psi \right\rangle&=\frac{\left\vert \Phi ^{+}\right\rangle_{12}}{\sqrt{2}}\Biggl[\Biggl(\alpha b \left\vert 1\right\rangle_3 + \beta a  \left\vert 0\right\rangle_3 \Biggl) \left\vert 0\right\rangle_4 +\alpha c \left\vert 0\right\rangle_3 \left\vert 1\right\rangle_4 \Biggl] \notag \\
&+\frac{\left\vert \Phi ^{-}\right\rangle_{12}}{\sqrt{2}}\Biggl[\Biggl(\alpha b \left\vert 1\right\rangle_3 - \beta a  \left\vert 0\right\rangle_3 \Biggl) \left\vert 0\right\rangle_4 +\alpha c \left\vert 0\right\rangle_3 \left\vert 1\right\rangle_4 \Biggl] \notag \\
&+\frac{\left\vert \psi ^{+}\right\rangle_{12}}{\sqrt{2}}\Biggl[\Biggl(\alpha a \left\vert 0\right\rangle_3 + \beta b  \left\vert 1\right\rangle_3 \Biggl) \left\vert 0\right\rangle_4 +\beta c \left\vert 0\right\rangle_3 \left\vert 1\right\rangle_4\Biggl] \notag \\
&+\frac{\left\vert \psi ^{-}\right\rangle_{12}}{\sqrt{2}}\Biggl[\Biggl(\alpha a \left\vert 0\right\rangle_3 - \beta b  \left\vert 1\right\rangle_3 \Biggl) \left\vert 0\right\rangle_4 -\beta c \left\vert 0\right\rangle_3 \left\vert 1\right\rangle_4 \Biggl]. \notag \\
\end{align}
Let us suppose that, after a measurement on qubits 1 and 2 in the EPR basis, Alice obtains $\left\vert \psi ^{+}\right\rangle$. In this case, the unnormalized state of qubits 3 and 4 is given by
\begin{equation}
\frac{1}{\sqrt{2}}\Biggl[\Biggl(\alpha a \left\vert 0\right\rangle_3 + \beta b  \left\vert a\right\rangle_3 \Biggl) \left\vert 0\right\rangle_4 +\beta c \left\vert 0\right\rangle_3 \left\vert 1\right\rangle_4 \Biggl].
\end{equation}
It's worth to note that the joint-state of the qubits 3 and 4 also contains EPR entanglement. This way, after a single qubit measurement, it is possible to complete the protocol.
If the qubit 4 it is projected on the $\left\vert 0\right\rangle$ state, then the state of qubit 3 (up to normalization) it will be $\left\vert \eta\right\rangle_3=\alpha a \left\vert 0\right\rangle_3 + \beta b  \left\vert 1\right\rangle_3.$ This state looks like the initial state $\ket{\phi}_1$. However, there is a imperfection due the coefficients $a$ and $b$. We can say that the initial state was teleported, but containing a imperfection. By other hand, if the qubit 4 it is projected on the state $\left\vert 1\right\rangle$, then there is no teleport. This way, we can conclude that the protocol works probabilistically. Another works investigating the using of W states as the quantum channel \cite{gorbachev2003can} were reported. In \cite{joo2003quantum}, the teleport of a single qubit state also was analyzed. Posteriorly, a important contribution was reported in \cite{PhysRevA.74.062320}, where was showed that there is type of W states that can be used for perfect teleportation and superdense coding. A generalization of that work it was reported in \cite{li2007states}.
In \cite{zhang2013quantum}, a composite W-Bell channel is used to execute quantum teleportation and superdense coding protocols.

\subsection{Teleportation of a W state}

\begin{figure}[h!]
\centering
\includegraphics[width=11.0 cm]{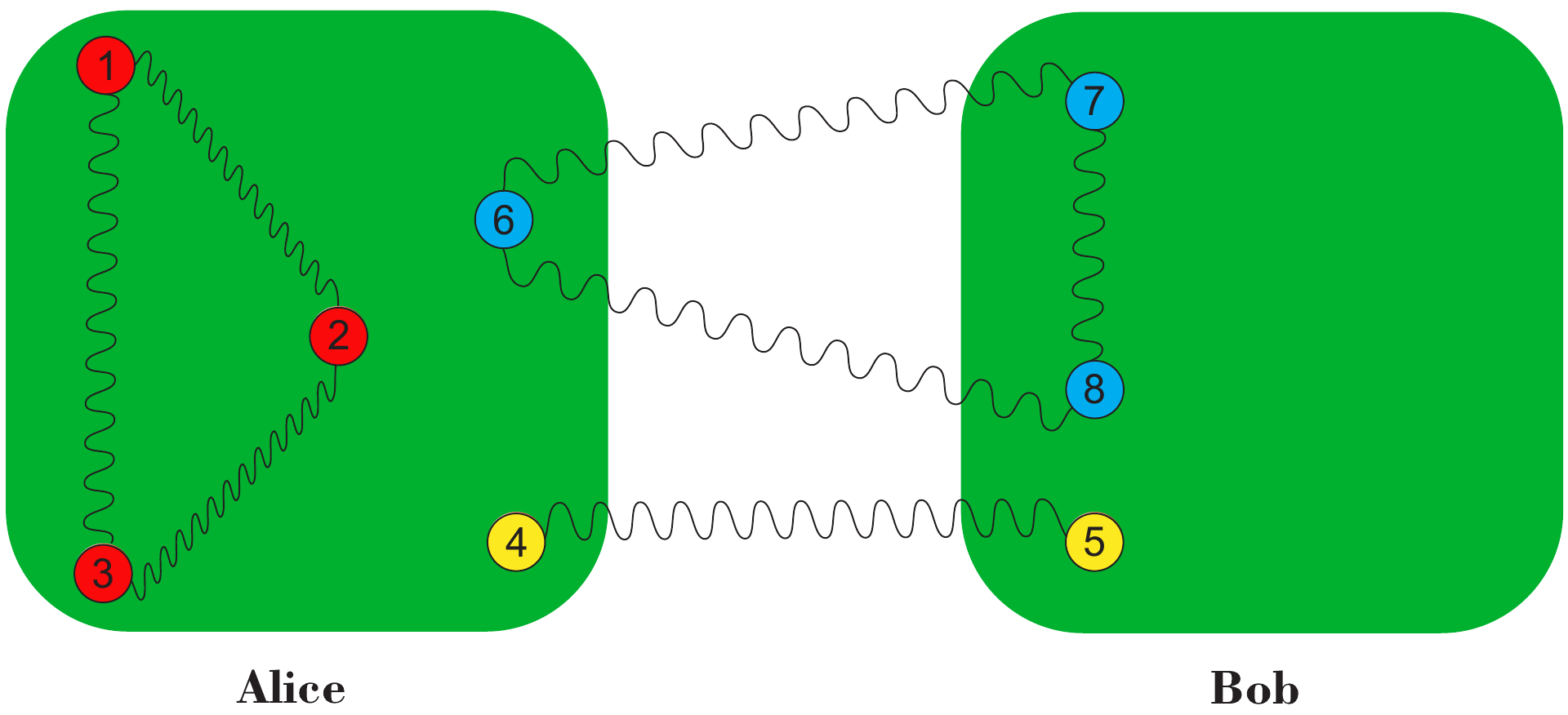}
\caption{A possible way to teleport a W state it is illustrated. A GHZ state and an EPR channel are used as the quantum channel.}
\label{teleporte_W}
\end{figure}
A quantum teleportation scheme of a tripartite W state it was presented in \cite{yi2002teleportation}. In that scheme, Alice wants to teleport the state
\begin{equation}
    \ket{\phi}_{123}=x\ket{001}_{123}+y\ket{010}_{123}+z\ket{100}_{123},
\end{equation}
on the particles 1, 2, 3. She shares two entangled states with Bob, a GHZ state and a EPR state, that will be used as the quantum channels. The EPR state is given by
\begin{equation}
 \ket{\psi}_{45}=a_0\ket{00}_{45}+a_1\ket{11}_{45},
\end{equation}
and the GHZ state is
\begin{equation}
    \ket{\eta}_{678}=c_0\ket{000}_{678}+c_1\ket{111}_{678}.
\end{equation}
The state of the whole system is
\begin{equation}
    \ket{\Psi}=\ket{\phi}_{123}\otimes \ket{\psi}_{45}\otimes \ket{\eta}_{678}.
\end{equation}
The particles 1, 2, 3, 4 and 6 are on the Alice's side, while particles 5, 7 and 8 are on Bob's side.
To execute the teleport, two EPR measurements are taken, on the particles 1, 6 and 3, 4. After the measurements, the resultant state involves the particles 2, 5, 7 and 8. Thus, it is necessary to trace out the particle 2. However, in order to complete the protocol, it is not make directly. Here, the protocol it is more complicated than the previous one we have discussed. Now, it is needed to employ combinations of Hadamard gate, CNOT gate and the introduction of a auxiliary particle. All procedures to do that are presented in the reference \cite{yi2002teleportation}.
More recently, another scheme to teleport a W state was reported in the reference \cite{gao2017teleportation}, where two W states are used as the quantum channel.

\subsection{Dense coding}

The original Dense Coding it was presented by using a two-entangled state as the resource, but
tripartite entanglement also it is suitable to make the protocol works \cite{doi:10.1002/qute.201900011}, and several schemes have been proposed. Two examples of controlled dense coding dealing with GHZ states can be accessed in \cite{hao2001controlled,huang2008controlled}. The W state also it is suitable to make a controlled dense coding protocol \cite{Yang2017}. A recent study about Super Dense Coding with W states can be accessed in \cite{zhou2018efficient}.
In \cite{ye2005scheme}, was proposed a scheme to implement Dense coding protocol by using tripartite entanglement in cavity QED. In \cite{roy2018deterministic} it was analyzed the use of GHZ and W states to make Deterministic Dense Coding.

\subsection{Quantum cryptography and Quantum Secure Communication}

Quantum Mechanics can also be used to perform cryptographic protocols. The basis of this field are founded by Wiensner \cite{wiesner1983conjugate}. Posteriorly, other fundamental steps were due to C. Bennet and G. Brassard (scheme known as BB84 protocol) \cite{bennett2014quantum} and A. Ekert \cite{ekert1991quantum}. After these initial developments, the use of Multipartite Entanglement has been considered.
In \cite{hillery1999quantum}, it was shown that a GHZ state can be used to make the Quantum Secret Sharing protocol. In \cite{joo2002quantum}, it was presented some variants of protocols of quantum secure communication dealing with W states.

Schemes to establish a three-party quantum secure communication using GHZ states are reported in \cite{jin2006three,man2006quantum}. In \cite{chen2008controlled}, it proposed a protocol for quantum secure direct communication utilizing W states.

\subsection{Other developments}

In \cite{christandl2005quantum}, it was presented  a quantum protocol to send and receive messages anonymously where $n$ players have a access a shared $n$-qubit GHZ entangles state.
Recently, it was reported a scheme to transmit a anonymous message in a network employing W-type states with N qubits \cite{lipinska2018anonymous}. In \cite{cruz2019efficient}, it were proposed quantum algorithms to the generation of GHZ  and W states of $n$ qubtis to be used on quantum networks.
In \cite{PhysRevA.83.012111}, tripartite entanglement it is examined in a noninertial frame when one of the parties is subjected to a uniform acceleration.
Tripartite entanglement also has been studied in the context of distillation of entanglement.

Methods of distillation of GHZ states  can be accessed in \cite{acin2000optimal,PhysRevA.65.024301}. Protocols for the optimal
distillation of W states were reported in \cite{yildiz2010optimal}.
In \cite{liao2014entanglement}, it is presented a method for entanglement purification of three-qubit states by using weak measurements.  In \cite{yuan2018one}, it was showed a distillation of GHZ-type states from two copies of a mixed state in a single step.

The relationship between tripartite entanglement and quantum computing also can be explored.
In \cite{gottesman1999demonstrating}, it was demonstrated that a GHZ state can be uses as an ingredient in the construction of a universal quantum computer.
The relationship between state complexity and quantum computing is discussed in \cite{doi:10.1002/andp.201400199}, including GHZ and W states in the analysis.
A discussion about the computational power of GHZ and W states can be accessed in \cite{D'Hondt:2006:CPW:2011665.2011668}.
In \cite{rajan2019quantum}, it was proposed a conceptual design for a quantum block-chain by using a temporal entangled GHZ state.

\section{Production of three-partite entanglement}

The production of multipartite entanglement have been attracted much attention. An experimental demonstration of five-qubits entanglement it was realized by Zhao and collaborators \cite{zhao2004experimental} in 2004. In 2011, Huang and colleagues have reported an experimental realization of an eight-photon GHZ state.
 In the same year, the creation of an entangled state of 14-qubit also it was presented \cite{monz201114}.
The first experimental production of of a three-photon GHZ state was reported in 1999 \cite{bouwmeester1999observation}. Recently, the deterministic generation of a 18-qubit entangled GHZ state it was achieved.
An experimental realization of a W state was reported in \cite{PhysRevLett.92.077901}.
In \cite{zeilinger1997three} it was reported a pioneer scheme for generating three-particle entanglements, using just two pairs of entangled particles from independent emissions.
Another pioneer work involving GHZ states it was due to Bouwmeester and colleagues \cite{bouwmeester1999observation}, observing the GHZ entanglement for three qubits in the polarization degree of freedom of photons. In \cite{laflamme1998nmr}, it was described the creation of maximally entangled GHZ states using Nuclear Magnetic Resonance (NMR). Since then, several works has been proposed, employing different frameworks.
A proposal for creation of GHZ and W states via quantum walks, for instance, can be accessed in \cite{ju2019creating}.
In \cite{erdosi2013proving}, it was presented a scheme to generate multipartite entanglement by using a single-neutron interferometer. In \cite{yamamoto2002polarization} was proposed a scheme to preparation of a W state using Parametric Down-Conversion.
Schemes to prepare  GHZ and W states of three distant atoms were reported in \cite{PhysRevA.75.044301}.  A method to generate a $n$-qubit W state in cavity QED it was reported in \cite{PhysRevA.73.014302}.
In \cite{gorbachev2003preparation}, a scheme to prepare W states from atomic ensembles it is presented.
The creation of GHZ and W states with a trapped-ion quantum computer was reported in \cite{roos2004control}.
The preparation of spin-qubit GHZ and W states in a quantum-dot-microcavity system it was discussed in \cite{kang2015effective}. It is also possible generate W states of three superconducting qubits \cite{kang2016fast}.
A scheme to generate entanglement between three atoms trapped in cavities via quantum Zeno dynamics it was proposed in \cite{chen2011tripartite}. In \cite{miry2015generation}, we can see a scheme to generate GHZ and W states from cavities with Jaynes–Cummings Hamiltonians.
In \cite{hamel2014direct}, it was showed a procedure to generate three-photon polarization entanglement, and is made the characterization of the produced states. Such states are used to make tests of local realism.
Recently, a scheme to generate GHZ states encoded on the path degree of freedom of three photons was presented in \cite{bernardo2018greenberger}. Another recent proposal for the generation of tripartite entanglement can be accessed in \cite{heo2019scheme}, where the creation of W states employing cross-Kerr nonlinearity and quantum dots it is addressed.
The production of larger states involving multi-qubits it is desirable to make protocols involving quantum networks \cite{mccutcheon2016experimental}. Several steps has been achieved in this direction.
A scalable scheme to create W states, for example, it was presented in 2017 \cite{PhysRevA.96.022319}.
In \cite{Yesilyurt:16}, it was presented a deterministic scheme for preparing W states of size of any power of 2.
A scheme to prepare a $N$-qubit GHZ state in a chain of four-level Rydberg atoms it was proposed in \cite{li2019adiabatic} recently. In \cite{omran2019generation}, it was demonstrate the creation of entangled states with up to 20 qubits.

\section{Detection and characterization of tripartite entanglement}

The detection of multipartite entanglement it is a hard task \cite{guhne2009entanglement}. Several efforts, theoretical and experimental, has been done to improve the methods to make it. The entanglement detection involves a key ingredient: the construction of Entanglement Witness \cite{bruss2002reflections,PhysRevX.8.031047} . In several practical situations, it is possible to certify entanglement through the violation of some inequality that indicates nonlocality. However, the number of inequalities increases when the number of qubits increases. While a measure of nonlocality depends on correlations, an Entanglement Witnesses it is an operator, corresponding to a physical observable.
The first GHZ state-analyzer was proposed in \cite{pan1998greenberger}, where two of the eight GHZ states can be distinguished using linear optical elements. After this first step, several contributions have been reported. In the \cite{qian2005universal}, it is presented a scheme for the universal tripartite GHZ-state analyzer using two-photon polarization QND\footnote{Here, QND means a Quantum Non-demolition, related with the idea of a QND measurement \cite{PhysRevA.73.012113}.} parity detectors based on the weak nonlinearity. The method allows discrimination of all the eight states with a probability near to the unity.
In \cite{wang2013nondestructive}, a method to construct a nondestructive n-qubit GHZ-state analyzer it is presented.
A GHZ state analyzer using only linear-optics elements through hyperentangled states\footnote{Hyperentanglement it is a type of entanglement dealing with multiple different degrees of freedom of a system. More details can be accessed in \cite{PhysRevA.72.052110,PhysRevA.91.062303}. } with polarization and momentum degrees of freedom can be viewed in \cite{song2013complete}.
Another scheme involving GHZ hyperentangled states it was recently proposed \cite{zheng2018self}.

In \cite{PhysRevA.81.032302}, it was investigated the existence of tripartite entanglement in a noninteracting Fermi gas, and some Entanglement Witness are introduced in that scenario.
A recent study about separability criteria of three-qubit GHZ state can be accessed in \cite{chen2015necessary}.
In \cite{li2017detection} are established sufficient conditions for detecting genuine tripartite entanglement, providing an operational point of view to measure and detect this type of entanglement.
Experimental schemes to identify the entanglement classes of tripartite states can be accessed in \cite{singh2018experimental,singh2018experimentally}.

A powerful tool to describe a density matrix of a system it is the method known as quantum state tomography, that allows the characterization of a quantum state \cite{PhysRevLett.105.150401}. These methods have been employed in the study of tripartite entangled states. In \cite{PhysRevLett.94.070402}, it was realized an experimental tomographic reconstruction of a three-photon polarization GHZ state. A more recent work using the method for GHZ states it is the reference \cite{lu2011characterization}.
In \cite{PhysRevLett.95.150404}, it was reported a scheme of experimental generation of a W state, and its full characterization by using quantum state tomography method.

\section{Remote preparation}

Besides the quantum teleportation protocol, another type of protocol dealing with the quantum state transfer it is the Remote Preparation Protocols (RSP) \cite{shi2018scheme}. Several possibilities of RSP can be done, involving different kinds of channels. It is possible to think in RSP to prepare single-qubit states, and also states with two or more qubits \cite{sun2019asymmetric}. Currently, this topic has been studied considering the preparation of multipartite entangled states. It is an interesting task since the production of entanglement remotely can help in using this resource to establish quantum communication between distant multi-users and establish quantum distributed computation \cite{kurpiers2018deterministic}.
In \cite{dai2006classical}, it was presented a scheme for remote preparation of a four-qubit GHZ state through two non-maximally entangled three-qubit GHZ states.
The remote preparation of a three-qubit state by using GHZ state it was considered in \cite{luo2010joint}. In \cite{zhan2013deterministic}, GHZ are used as the quantum channel for the remote preparation of arbitrary states of one and two qubits. A scheme for preparing atomic states remotely by using GHZ states it is also possible \cite{khosa2010remote}.
In \cite{an2010joint}, W states are used for remote preparation of single and two-qubit states.
In \cite{choudhury2015joint}, two maximally entangled GHZ states are used as the quantum channel for the remote preparation of an arbitrary equatorial two-qubit state.
A scheme for remote preparation of a two-qubit state by using two W-type states as the quantum channel also it was reported \cite{wang2012classical}.
The remote preparation of $m$-qubit states by using non-maximally GHZ entangled state it was studied in \cite{wang2015generalized}.
In \cite{PhysRevA.94.042329}, it was reported an experimental demonstration of remote preparation of three-photon entangled states by using a single photon measurement.
In \cite{lv2018multiparty}, it was presented a scheme to joint remote preparation of a $m$-qudit state by using a $d$-dimensional GHZ state.
In \cite{xia2012deterministic}, EPR pairs are used as the quantum channel to the preparation of an arbitrary three-qubit state.
Tripartite entangled states not only can be used to remote preparation of single and two-qubit states, but they also can be themselves remotely prepared. In \cite{luo2010deterministic}, it was presented a scheme for preparing W states remotely by using two four-particle GHZ states as the quantum channel.
A proposal to remote preparation of a three-particle state employing a three-particle orthonormal basis projective measurement can be accessed in \cite{peng2008scheme}.
An idea for remotely preparing W states of three and four-qubit utilizing tripartite GHZ states it was presented in \cite{wang2015efficient}.
In \cite{wu2018highly}, it was reported an efficient scheme for remote preparation of a $2^n$-qubit W state via $n$ three-qubit GHZ states.
A scheme to prepare a tripartite equatorial state by using three maximally entangled GHZ states it was proposed in \cite{sang2019deterministic}.
In \cite{choudhury2019remote}, it was presented a protocol to remotely preparing a tripartite W-type state by using an eight-qubit state as the quantum resource.

\section{Continuous-variable systems}

Until this point, we have explored several aspects involving tripartite entanglement of discrete variables, that refers to qubit-based developments, like photon polarization or the spin of electron. It is worth to note that, it is also possible dealing with multipartite entangled states  in the domain of continuous variables \cite{eisert2003introduction,adesso2007entanglement}. Several quantum information protocols has been considered in that domain, where the variables has a continuous spectrum of eigenvalues. For instance, an analysis of quantum teleportation of quantum states with continuous variables was presented in \cite{PhysRevLett.80.869}. In \cite{PhysRevA.67.032302}, an experimental investigation of this subject was reported. The entanglement swapping involving continuous variables was investigated in \cite{PhysRevLett.83.2095} and \cite{PhysRevLett.114.100501}.
A study approaching necessary conditions of separability of a multiparty continuous-variable state can be viewed in \cite{PhysRevA.67.052315}. A detailed discussion about quantum information and continuous variables it is presented in \cite{adesso2014continuous}.
A complete analysis of  three-mode Gaussian  entangled states it is showed in \cite{PhysRevA.73.032345}. In that work, the counterparts of GHZ and W states in the continuous-variable scenario are presented.
Experimentally, the creation of a tripartite entangled state in continuous-variables it is reality \cite{aoki2003experimental}.
In \cite{coelho2009three}, it was demonstrated the generation of entanglement among three beams of light with different wavelengths.
In \cite{shalm2013three}, there is a experimental demonstration of tripartite entanglement where correlations involving energies and emission times of photons takes place.
In \cite{PhysRevA.90.052321}, there is a discussion about hierarchies of separability criteria of multipartite entanglement for continuous-variable states.

\section{Noisy environments}

If we desire to consider more realistic scenarios to generation, detection and the protocols involving entanglement, we need to include some effects of the environment.
Because the interaction of the system of interest and the environment, several errors can affect the quality of the quantum protocols. The effect of decoherence, for example, can degrade the entanglement of quantum channels. Thus, several works has been proposed to analyze the noise effect in the execution of quantum protocols. Examples of noisy quantum teleportation protocols can be accessed in \cite{liang2013quantum,PhysRevA.93.062330}.

The effect of noise in tasks dealing with tripartite entanglement also has been attracting attention. The influence of bit-flip noise, for example, in the entanglement and non-locality of GHZ states it was investigated in  a recent work \cite{zhao2018tripartite}. In \cite{PhysRevLett.100.080501}, it was investigated the decay of entanglement of $N$-particle GHZ states due the interaction with the environment.
The impact of a decoherence process in the entanglement of GHZ and W states it was analyzed in \cite{kenfack2018decoherence}.
 In \cite{guo2012quantum}, it was studied the quantum discord of a W state in the presence of noise. The evolution of the quantum discord under noisy effects to GHZ and W states it is analyzed in \cite{mahdian2012quantum}. In \cite{metwally2015dynamics}, some properties of GHZ and W states under a depolarizing noise are discussed. In \cite{lionel2017effects}, it was studied noise effects on the quantum correlations of a three-qubits system.
A comparison between GHZ and W noisy-channels to execute the quantum teleport of a single qubit state can be accessed in the references \cite{PhysRevA.78.012312,hu2011robustness}.
In \cite{metwally2014entanglement}, it was investigated the effect of the generalized amplitude damping channel on GHZ states.
In \cite{chun2010controlled}, it was addressed the problem of teleporting a unknown atomic state through a noisy GHZ channel.
The entanglement of GHZ states with decoherence in non-inertial frames it is discussed in \cite{khan2012entanglement}. In \cite{zounia2017quantum} it was studied the quantum teleport by using noisy bipartite and tripartite entangled states which one of the users experiments a uniform acceleration.

In \cite{yang2019three},  schemes based on GHZ states are presented to make quantum secret sharing protocols immune to the some kinds of collective noise.
The effects of some kind of noises in GHZ channels on  remote preparation of states also has been investigated, for a single qubit state \cite{liang2011remote,liang2015effects} and also for a two-qubit state \cite{zhong2013deterministic}.
In \cite{zhang2018cyclic}, GHZ states are considered for remote state preparation of quantum states in noisy environments.
The relationship between weak measurement and GHZ entanglement distribution in the presence of noise also it was investigated \cite{wang2016effect}.
The robustness of GHZ and W states against decoherence it was experimentally investigated in \cite{PhysRevA.97.022302}.
In \cite{jiang2019cyclic}, it was presented a scheme for quantum communication in noisy environments by using a hybrid channel Bell-GHZ.
The teleport of a GHZ state in the presence of noisy channels it was investigated in \cite{cunha2017non}.
An investigation of the robustness of cat-like states under Pauli noises it was reported in \cite{wang2019improving}.

\section{Conclusions}

In this work, we have addressed several aspects of tripartite entanglement. We have started revisiting the main aspects of this type of entanglement. We also have done some remarks about bipartite entanglement to clarify our discussion about tripartite entanglement. Then, we explore in more details the two inequivalent classes of tripartite entanglement, namely GHZ and W, and define the corresponding basis. We keep forward making a revision about the relationship between Bell's Theorem and the GHZ states. Posteriorly, we gave some examples of quantum information protocols working with tripartite entanglement. We give special attention to quantum teleportation protocols, making some calculations and exhibiting the corresponding schemes on several figures. After this, we list the main contributions in the literature relative to the production and detection of tripartite entanglement. Then, we explore some aspects relative to Remote Preparation protocols. We also make some aspects of tripartite entanglement dealing with tripartite entanglement. Finally, we have discussed the study of tripartite entanglement in several scenarios in the presence of a noisy environment.\\\\\\

\textit{This work was partially supported by the Brazilian agencies CAPES, CNPq and FAPEMA. EOS acknowledges CNPq Grants
427214/2016-5 and 303774/2016-9, and FAPEMA Grants 01852/14 and 01202/16. MMC acknowledges CAPES Grant 88887.358036/2019-00.
We thanks to M.G.M Moreno for useful suggestions.}

\bibliography{References}

\newpage

\appendix

\section*{\textbf{ERRATUM: Tripartite entanglement: Foundations and applications}}

On section V, subsection A of this manuscript,  we have discussed the quantum teleportation of a single qubit by using a GHZ channel and EPR measurements. We explore some details of the protocol presented in \cite{karlsson1998quantum}, and review it using a different notation. Even though equations (47) and (50) are correct,  equation (41) is not. Moreover, the tables I, II and III in the manuscript are not correct. The main goal of the present erratum is to present an updated version including corrections of the mistakes above mentioned.

 Under the presented scheme, Alice, Bob, and Charlie share a GHZ state, as shown in figure \ref{canal_tripartite}. Moreover, Alice  has a separated qubit (qubit 1), whose state she intends to teleport. The state of the whole system is
\begin{equation}
\ket{\Psi}=[\alpha\ket{0}+\beta\ket{1}]_1 \otimes \frac{1}{\sqrt{2}}[\ket{000}+\ket{111}]_{234}=\sum_{i=0}^1 \alpha_i \ket{i}_1 \otimes \sum_{j=0}^1 \frac{1}{\sqrt{2}} \ket{jjj}_{234}.  
\end{equation}

In this protocol Alice carries out a Bell-measurement on  qubits 1 and 2, then it is appropriate to write the state as
\begin{equation}
\ket{\Psi}=\frac{1}{\sqrt{2}}\sum_{i,j} \alpha_i \ket{ij}_{12} \ket{jj}_{34}.
\end{equation}

Let us calculate the projection of $\ket{\Psi}$ into the $m,n$ element of the EPR basis on qubits 1 and 2, $\bra{\phi_{mn}^{(2)}}_{12}$ :
\begin{equation}
     \bra{\phi_{mn}}_{12}\hspace{0.05cm}\ket{\Psi}=\ket{\eta}_{34}=\frac{1}{2}\sum_{ijk}\alpha_i (-1)^{mk}\left\langle k,k \oplus n \vert i,j \right\rangle_{12}\otimes \ket{j,j}_{34},
\end{equation}
\begin{equation}
     \bra{\phi_{mn}}_{12}\hspace{0.05cm}\ket{\Psi}=\ket{\eta}_{34}=\frac{1}{2}\sum_{ijk}\alpha_i (-1)^{mk} \delta_{k,i}\delta_{k \oplus n,j} \ket{j,j}_{34}=\frac{1}{2}\sum_{k}\alpha_k (-1)^{mk}  \ket{k \oplus n,k \oplus n}_{34},
\end{equation}
\begin{equation}
     \bra{\phi_{mn}}_{12}\hspace{0.05cm}\ket{\Psi}=\ket{\eta}_{34}=\frac{1}{2}\sum_{k}\alpha_k (-1)^{mk}  \ket{k \oplus n,k \oplus n}_{34}.
\end{equation}
 As result, Charlie and Bob share an EPR-like state:
\begin{equation}
    \ket{\eta}_{34}=\ket{\eta_{mn}}_{34}=\sum_{k=0}^1 (-1)^{mk}\alpha_k \ket{k \oplus n, k \oplus n}_{34}.
\end{equation}
 Thus, Charlie and Bob  share an imperfect EPR state (with coefficients $\alpha_0$ and $\alpha_1$) which depends on Alice's measurement outcome. 
For example, when Alice obtains the state $\ket{\phi_{00}},$  Bob and Charlie share the state
\begin{equation}
\ket{\eta_{00}}_{34}=\sum_k \alpha_k \ket{k,k}=\alpha_0 \ket{00}+\alpha_1 \ket{11}.
\label{eta00_eq}
\end{equation}
\begin{table}[h]
\caption{Results after Alice's measurement and states shared by Bob and Charlie.}
\centering
\begin{tabular}{ccc}
\toprule Alice's result & & State shared by Bob and Charlie\\
\midrule $m$ & $n$ & state \\
\midrule 0 & 0 & $\ket{\eta_{00}}= \alpha_0\ket{00}+ \alpha_1\ket{11} $ \\
\midrule 0 & 1 & $\ket{\eta_{01}}=\alpha_0\ket{11}+ \alpha_1\ket{00} $ \\
\midrule 1 & 0 & $\ket{\eta_{10}}=\alpha_0\ket{00}- \alpha_1\ket{11} $ \\
\midrule 1 & 1 & $\ket{\eta_{11}}=\alpha_0\ket{11}-\alpha_1 \ket{00} $ \\
\bottomrule
\end{tabular}
\label{eta}
\end{table}
 Table \ref{eta} exhibits the states shared by Bob and Charlie after Alice's measurement. Let us explore in more detail the case $\ket{\eta_{00}}.$ As mentioned in \cite{karlsson1998quantum}, to proceed with the protocol, Bob or Charlie should carry out a measurement on a single qubit. Let us suppose that Alice wants to teleport the state of her qubit to Charlie. Then, she asks Bob to perform a measurement on his qubit. For these purposes, Bob uses the following basis:
\begin{equation}
\ket{0}_3=\sin\theta \ket{x_0}_3+\cos\theta \ket{x_1}_3, \;\;\;\;\;
\ket{1}_3=\cos\theta \ket{x_0}_3-\sin\theta \ket{x_1}_3.
\end{equation}
Substituting these expressions on equation (\ref{eta00_eq}), we have:
\begin{equation}
\ket{\eta_{00}}_{34}=\alpha_0 (\sin\theta\ket{x_1}_3+\cos\theta\ket{x_2}_3)\ket{0}_4 +\alpha_1 (\cos\theta\ket{x_1}_3-\sin\theta\ket{x_2}_3)\ket{0}_4.
\end{equation}
We can also write this expression as
\begin{equation}
\ket{\eta_{00}}_{34}=\ket{x_1}_3(\alpha_0 \sin\theta\ket{0}_4+\alpha_1\cos\theta\ket{1}_4)+\ket{x_2}_3(\alpha_0\cos\theta \ket{0}_4-\alpha_1\sin\theta \ket{1}_4).
\end{equation}
Back to the general case, by defining $b_0 \equiv \sin\theta$ and $b_1\equiv\cos\theta$, we can then write Bob's measurement basis as:
\begin{equation}
\ket{k}_3=\sum_{j=0}^1(-1)^{j(k\oplus n)}b_{j \oplus k \oplus n}\ket{x_j}_3.    
\end{equation}
\begin{equation}
\ket{\eta_{mn}}_{34}=\sum_{k=0}^1 (-1)^{mk}\alpha_k \left(\sum_{j=0}^1(-1)^{j(k\oplus n)}b_{j \oplus k \oplus n}\ket{x_j}_3   \right)\ket{k \oplus n}_4.
\end{equation}
To obtain the final state of Charlie's qubit, we can calculate the projection of $\ket{x_{j}}$ on $\ket{\eta_{mn}}_{34}$, where $j=0,1.$
\begin{equation}
 \ket{\chi_{mnj}}_4=\bra{x_j}_3 \hspace{0.05cm}\ket{\eta_{mn}}_{34}.
\end{equation}
Table \ref{eta} shows all possibilities after Alice's and Bob's measurements.
\begin{table}[]
\caption{Alice's and Bob's results, and  final state obtained by Charlie.}
\centering
\begin{tabular}{cccc}
\toprule Alice's result & & Bob's result   & State (unnormalized) of  Charlie's qubit\\
\midrule $m$ & $n$ & $\ket{x_i}$  & state \\
\midrule 0 & 0 & $\ket{x_0}$  & $ \alpha_0\sin\theta\ket{0}+ \alpha_1\cos\theta\ket{1} $ \\
\midrule 0 & 0 & $\ket{x_1}$ & $\alpha_0\cos\theta\ket{0}- \alpha_1\sin\theta\ket{1} $ \\
\midrule 0 & 1 & $\ket{x_0}$ & $\alpha_0\cos\theta\ket{1}+ \alpha_1\sin\theta\ket{0} $ \\
\midrule 0 & 1 & $\ket{x_1}$ & $-\alpha_0\sin\theta\ket{1}+\alpha_1\cos\theta \ket{0} $ \\
\midrule 1 & 0 & $\ket{x_0}$ & $\alpha_0\sin\theta\ket{0}-\alpha_1\cos\theta \ket{1} $ \\
\midrule 1 & 0 & $\ket{x_1}$ & $\alpha_0\cos\theta\ket{0}+\alpha_1\sin\theta \ket{1} $ \\
\midrule 1 & 1 & $\ket{x_0}$ & $\alpha_0\cos\theta\ket{1}-\alpha_1\sin\theta \ket{0} $ \\
\midrule 1 & 1 & $\ket{x_1}$ & $-\alpha_0\sin\theta\ket{1}-\alpha_1\cos\theta \ket{0} $ \\
\bottomrule
\end{tabular}
\label{mnx}
\end{table}
\begin{table}
\caption{Alice's and Bob's results \textcolor{blue}{,} and corresponding operations needed to recover the desired state.}
\centering
\begin{tabular}{cc}
\toprule  Result: \hspace{0.05cm} $(m,n,i)$ & Operation \\
\midrule $(0,0,0)$ & $\hat{I}$  \\
\midrule $(0,0,1)$ & $\hat{\sigma}_z$  \\
\midrule $(0,1,0)$ & $\hat{\sigma}_x$  \\
\midrule $(0,1,1)$ & $\hat{\sigma}_x\hat{\sigma}_z$  \\
\midrule $(1,0,0)$ & $\hat{\sigma}_z$ \\
\midrule $(1,0,1)$ & $\hat{I}$ \\
\midrule $(1,1,0)$ & $\hat{\sigma}_z\hat{\sigma}_x$  \\
\midrule $(1,1,1)$ & $\hat{\sigma}_x$ \\
\bottomrule
\end{tabular}
\label{unitarias1}
\end{table}

The indexes $m,n,j$ are related to the measurement outcomes on qubits possessed by Alice and Bob.  In order to recover the desired state, Charlie has to apply a unitary transformation on his qubit. Which unitary he will use, depends on the values the indexes $(m,n,i)$ take.  Table \ref{unitarias1} lists the corresponding operation for each result, i.e. for the case $(0,0,0)$, the state (up to normalization) is
\begin{equation}
\ket{\chi_{000}}=\alpha_0 \sin\theta\ket{0}_4+\alpha_1\cos\theta \ket{1}_4.
\end{equation}
Thus, the operation is the identity. The normalized state is
\begin{equation}
\ket{\chi_{000}}_f = \frac{\alpha_0 \sin\theta\ket{0}_4+\alpha_1\cos\theta \ket{1}_4}{\sqrt{|\alpha_0|^2\sin^2\theta +|\alpha_1|^2\cos^2\theta}}.
\end{equation}
All results are summarized in tables \ref{eta} and \ref{unitarias1}.

\end{document}